\documentclass[12pt,draftcls,onecolumn] {IEEEtran}
\usepackage[cmex10]{amsmath}
\usepackage{pifont}
\usepackage{stmaryrd}
\usepackage{mathrsfs}
\usepackage{amsmath}
\usepackage{amssymb}
\usepackage{multirow}
\usepackage{graphicx}
\usepackage{subfigure}
\usepackage{color}
\usepackage{subeqnarray}
\usepackage{slashbox}
\usepackage{multirow}
\usepackage{cases}
\usepackage[colorlinks,
            linkcolor=black,
            anchorcolor=black,
            citecolor=black
            ]{hyperref}

\newcommand{\bs}{{\mbox{\boldmath{$s$}}}}
\newcommand{\bu}{{\mbox{\boldmath{$u$}}}}

\newcommand{\bw}{{\mbox{\boldmath{$w$}}}}

\newcommand{\bD}{{\mbox{\boldmath{$D$}}}}

\newcommand{\bH}{{\mbox{\boldmath{$H$}}}}

\newcommand{\bS}{{\mbox{\boldmath{$S$}}}}

\newcommand{\bW}{{\mbox{\boldmath{$W$}}}}

\newcommand{\bgamma}{{\mbox{\boldmath{$\gamma$}}}}

\begin{document}
\title{IRCI Free Range Reconstruction for SAR Imaging with Arbitrary Length OFDM Pulse}
\author{Tian-Xian Zhang, Xiang-Gen Xia, and Lingjiang Kong
\thanks{Tian-Xian Zhang and Lingjiang Kong are with the School of
Electronic Engineering, University of Electronic Science and Technology of
China, Chengdu, Sichuan, P.R. China, 611731. Fax: +86-028-61830064, Tel:
+86-028-61830768, E-mail: tianxian.zhang@gmail.com, lingjiang.kong@gmail.com.
Zhang's research was supported by the Fundamental Research Funds for the
Central Universities under Grant ZYGX2012YB008 and by the China Scholarship
Council (CSC) and was done when he was visiting the University of Delaware,
Newark, DE 19716, USA. Xiang-Gen Xia is with the Department of Electrical and
Computer Engineering, University of Delaware, Newark, DE 19716, USA. Email:
xxia@ee.udel.edu. Xia's research was partially supported by the Air Force
Office of Scientific Research (AFOSR) under Grant FA9550-12-1-0055. }}

\maketitle

\vspace{-0.18in}

\begin{abstract}
Our previously proposed OFDM with sufficient cyclic prefix (CP) synthetic
aperture radar (SAR) imaging algorithm is inter-range-cell interference (IRCI)
free and achieves ideally zero range sidelobes for range reconstruction. In
this OFDM SAR imaging algorithm, the minimum required CP length is almost equal
to the number of range cells in a swath, while the number of subcarriers of an
OFDM signal needs to be more than the CP length. This makes the length of a
transmitted OFDM sequence at least almost twice of the number of range cells in
a swath and for a wide swath imaging, the transmitted OFDM pulse length becomes
long, which may cause problems in some radar applications. In this paper, we
propose a CP based OFDM SAR imaging with arbitrary pulse length, which has IRCI
free range reconstruction and its pulse length is independent of a swath width.
We then present a novel design method for our proposed arbitrary length OFDM
pulses. Simulation results are presented to illustrate the performances of the
OFDM pulse design and the arbitrary pulse length CP based OFDM SAR imaging.
 \\
 \\
{\bf EDICS}: RAS-SARI (Synthetic aperture radar/sonar and imaging), RAS-IMFR
(Radar image formation and reconstruction).
\end{abstract}


\maketitle

\clearpage
\section{Introduction}\label{Introduction}
Orthogonal frequency-division multiplexing (OFDM) signals are firstly presented
for radar signal processing in \cite{LevanonIEEPMultiF2000276}, and recently
studied and used in radar applications, such as moving target detection
\cite{SenSPLOFDMDetection2009592, SenTSPOFDMmultipath201178, SenTGRSPAPR20131},
low-grazing angle target tracking \cite{SenTSPOFDMMIMO20103152} and
ultrawideband (UWB) radar applications \cite{GarmatyukWidebandOFDM20091}. The
common OFDM signals for digital communications, such as the digital audio
broadcast (DAB), digital video broadcast (DVB), Wireless Fidelity (WiFi) or
worldwide inoperability for microwave access (WiMAX) signals, are also
investigated for radar applications in \cite{BergerSTSPOFDMPassive2010226,
ColoneAESOFDMPassive2011240, FalconeExperimental2010516,
ColoneAmbiguityFunction2010689, ChettyPassiveWiMAX2010188, QingWiMAX20091}.
Using OFDM signals for synthetic aperture radar (SAR) applications is proposed
in \cite{RicheOFDMSAR122156, RicheOFDMSAR12278, GarmatyukOFDMSAR06237,
GarmatyukGRSOFDMSAR20113780, GarmatyukGRSLOFDMSAR2012808,
GutierrezAESOFDMPassiveSAR2013945}. In \cite{RicheOFDMSAR122156,
RicheOFDMSAR12278}, an adaptive OFDM signal design is studied for range
ambiguity suppression in SAR imaging. The reconstruction of cross-range
profiles is studied in \cite{GarmatyukGRSOFDMSAR20113780,
GarmatyukGRSLOFDMSAR2012808}. However, all the existing OFDM SAR signal
processing algorithms have not considered the feature of OFDM signals with
sufficient cyclic prefix (CP) as used in communications systems. In
\cite{TxzOFDMSAR}, we have proposed a sufficient CP based OFDM SAR imaging
algorithm. By using a sufficient CP, the inter-range-cell interference (IRCI)
free and ideally zero range sidelobes for range reconstruction can be obtained,
which provides an opportunity for high range resolution SAR imaging. On the
other hand, according to our analysis, the CP length, the transmitted OFDM
pulse length and the minimum radar range need to be increased with the increase
of a swath width, since the sufficient CP length is almost equal to the number
of range cells in a swath, while the number of subcarriers of the OFDM signal
needs to be more than the CP length. Then, the transmitted OFDM sequence with
sufficient CP should be at least almost twice of the number of range cells in a
swath. Meanwhile, the CP sequence needs to be removed at the receiver to
achieve the IRCI free range reconstruction. Thus, this algorithm may need a
long transmitted pulse and suffer high transmitted energy redundancy in case of
wide swath imaging, which may cause problems in some radar applications.

Although OFDM signals have been widely used in practical digital communications
and studied for radar applications, the potential high peak-to-average power
ratio (PAPR) of OFDM signals may cause problems for communications applications
\cite{SeungIEEEWCPAPRoverview200556} and radar applications
\cite{SenTSPOFDMmultipath201178}, because the envelope of OFDM signals is
time-varying. In power amplifier of the transmitter, a constant envelope
waveform can be magnified efficiently in the saturation region. However, the
amplifier should be operated in the limited linear region for a time-varying
signal to avoid causing nonlinear distortion. Many PAPR reduction techniques
have been studied as, for example, in \cite{prasad2004ofdm}.

In this paper, we propose a sufficient CP based OFDM SAR imaging with arbitrary
pulse length that is independent of a swath width. Firstly, we establish the
arbitrary pulse length OFDM SAR imaging system model by considering the feature
of OFDM signals with sufficient CP, where the CP part is all zero. We then
derive a sufficient CP based range reconstruction algorithm with an OFDM pulse,
whose length is independent of a swath width. To investigate the
signal-to-noise ratio (SNR) degradation caused by the range reconstruction, we
also analyze the change of noise power in every step of the range
reconstruction. By considering the PAPR of the transmitted OFDM pulses and the
SNR degradation within the range reconstruction, we propose a new OFDM pulse
design method. We then present some simulations to demonstrate the performance
of the proposed OFDM pulse design method. By comparing with the range Doppler
algorithm (RDA) SAR imaging method using LFM signals, we present some
simulations to illustrate the performance of the proposed the arbitrary pulse
length OFDM SAR imaging algorithm. We find that, with a designed arbitrary
length OFDM pulse from our proposed method, this algorithm can still maintain
the advantage of IRCI free range reconstruction with insignificant SNR
degradation and completely avoid the energy redundancy.

The remainder of this paper is organized as follows. In Section
\ref{Problem_For}, we briefly recall the CP based OFDM SAR algorithm proposed
in \cite{TxzOFDMSAR} and describe the problem of interest. In Section
\ref{APLength}, we propose CP based arbitrary pulse length OFDM SAR. In Section
\ref{sec4}, we propose a new arbitrary length OFDM sequence design algorithm.
In Section \ref{Simulation}, we show some simulation results. Finally, in
Section \ref{Conclusion}, we conclude this paper.

\section{CP Based OFDM SAR and Problem Formulation} \label{Problem_For}
\begin{figure}[t]
\begin{center}
\includegraphics[width=0.6\columnwidth,draft=false]{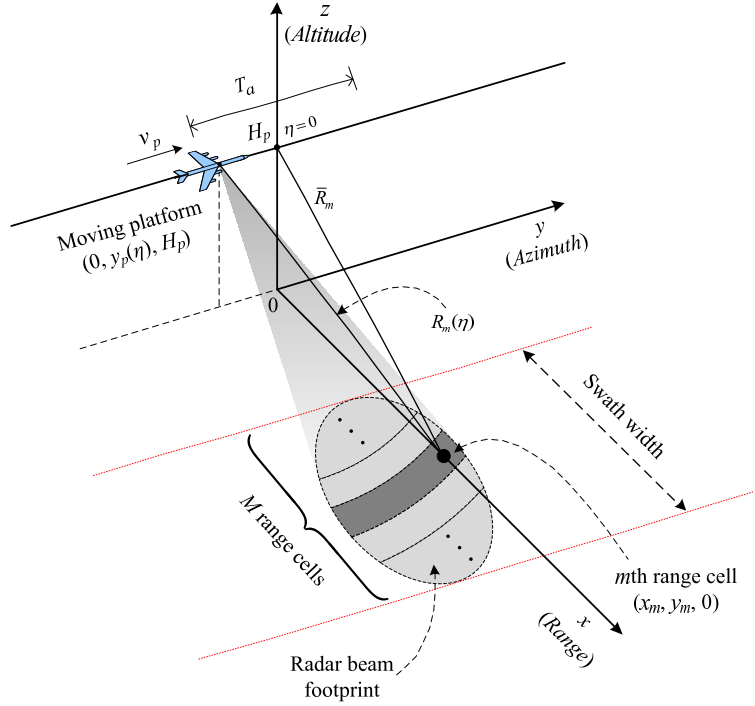}
\end{center}
\caption{Monostatic stripmap SAR geometry.}\label{geometry}
\end{figure}

In this section, we first briefly recall the CP based OFDM SAR model proposed
in \cite{TxzOFDMSAR} and then see its required pulse length problem. Consider
the monostatic broadside stripmap SAR geometry as shown in Fig. \ref{geometry}.
The radar platform is moving parallelly to the $y$-axis with an instantaneous
coordinate $(0, y_p(\eta), H_p)$, $H_p$ is the altitude of the radar platform,
$\eta$ is the relative azimuth time referenced to the time of zero Doppler,
$T_a$ is the synthetic aperture time defined by the azimuth time extent the
target stays in the antenna beam. For convenience, let us choose the azimuth
time origin $\eta=0$ to be the zero Doppler sample. Consider an OFDM signal
with $N$ subcarriers, a bandwidth of $B$ Hz, and let
$\bS=\left[S_0,S_1,\ldots,S_{N-1}\right]^T$ represent the complex weights
transmitted over the subcarriers, $(\cdot)^T$ denotes the transpose, and
$\sum_{i=0}^{N-1}\left|S_i\right|^2=1$. Note that, although this sequence $S_i$
is rather general, in \cite{TxzOFDMSAR}, a pseudo random sequence $S_i$ with
constant module is proposed to be used for achieving the optimal SNR at the
receiver. Then, a discrete time OFDM signal is the inverse fast Fourier
transform (IFFT) of the vector $\bS$ and the corresponding time domain OFDM
signal is
\begin{equation}\label{OFDM}
s(t)=\frac{1}{\sqrt{N}}\sum_{k=0}^{N-1}S_k
\textrm{exp}\left\{j2\pi k\Delta ft\right\},\ t\in \left[0,T+T_{GI}\right],
\end{equation}
where $\Delta f=\frac{B}{N}=\frac{1}{T}$ is the subcarrier spacing.
$\left[0,T_{GI}\right)$ is the time duration of the guard interval that
corresponds to the CP in the discrete time domain as we shall see later in more
details and its length $T_{GI}$ will be specified later too, $T$ is the length
of the OFDM signal excluding CP. Due to the periodicity of the exponential
function $\textrm{exp}(\cdot)$ in (\ref{OFDM}), the tail part of $s(t)$ for $t$
in $\left(T,T+T_{GI}\right]$ is the same as the head part of $s(t)$ for $t$ in
$\left[0,T_{GI}\right)$.

After the demodulation to baseband, the complex envelope of the received signal
from all the range cells in the swath can be written in terms of fast time $t$
and slow time $\eta$
\begin{equation}\label{Receive}
u(t,\eta)=\frac{1}{\sqrt{N}}\sum\limits_{m}{g_m}\textrm{exp} \left\{-j4\pi
f_c\frac{R_m(\eta)}{c}\right\}\sum\limits_{k=0}\limits^{N-1}S_k
\textrm{exp}\left\{\frac{j2\pi k}{T}
\left[t-\frac{2R_m(\eta)}{c}\right]\right\}+w(t,\eta),
\end{equation}
where $f_c$ is the carrier frequency, $g_m$ is the radar cross section (RCS)
coefficient caused from the scatterers in the $m$th range cell within the radar
beam footprint, and $c$ is the speed of light. $w(t,\eta)$ represents the
noise. $R_m(\eta)=\sqrt{\bar{R}_m^2+v_p^2\eta^2}$ is the instantaneous slant
range between the radar and the $m$th range cell with the coordinate $(x_m,
y_m, 0)$, $\bar{R}_m=\sqrt{x_m^2+H_p^2}$ is the slant range when the radar
platform and the target in the $m$th range cell are the closest approach, and
$v_p$ is the effective velocity of the radar platform.

At the receiver, the received signal is sampled by the A/D converter with
sampling interval length $T_s=\frac{1}{B}$ and the range resolution is
$\rho_r=\frac{c}{2B}=\frac{c}{2}T_s$. Assume that the swath width for the radar
is $R_w$. Then, a range profile can be divided into $M=\frac{R_w}{\rho_r}$
range cells that is determined by the radar system. According to the analysis
in \cite{TxzOFDMSAR}, $M$ range cells correspond to $M$ paths in
communications, which include one main path (i.e., the nearest range cell) and
$M-1$ multipaths. In order to avoid the IRCI (corresponding to the intersymbol
interference (ISI) in communications) between different range cells, the CP
length should be at least equal to the number of multipaths ($M-1$). For
convenience, we set CP length as $M-1$ in \cite{TxzOFDMSAR}, and then the guard
interval length in (\ref{OFDM}) is $T_{GI}=(M-1)T_s$. Notice that $T=NT_s$.
Thus, the time duration of an OFDM pulse is $T_o=T+T_{GI}=(N+M-1)T_s$.
Meanwhile, to completely avoid the IRCI between different range cells, the
number, $N$, of the OFDM signal subcarriers should satisfy $N\geq M$ as we have
analyzed in \cite{TxzOFDMSAR} and also well understood in communications
applications \cite{prasad2004ofdm}. Therefore, the transmitted pulse duration
$T_o$ is increased with the increase of the swath width. For example, if we
want to increase the swath width to $10$ km, the transmitted pulse duration
$T_o$ should be increased to about $133.3\ \mu$s. The pulse length here is much
longer than the traditional radar pulse, which might be a problem, especially,
for covert/military radar applications. Therefore, it is important to achieve
OFDM SAR imaging with arbitrary pulse length that is independent of a swath
width, and in the meantime it also has ideally zero IRCI. This is the goal of
the remainder of this paper.

\section{CP Based Arbitrary Pulse Length OFDM SAR}\label{APLength}
The main idea of the following study is to generate a pulse $s(t),\ t\in
\left[0,T+T_{GI}\right]$,  such that $s(t)=0$ for $t\in [0,T_{GI})$ and also
for $t\in \left(T,T+T_{GI}\right]$ with an arbitrary $T$ for $T>T_{GI}$, and
$s(t)$ is an OFDM signal in (\ref{OFDM}) for $t\in \left[T_{GI},T\right]$.
However, if the non-zero segment $s(t)$ for $t\in \left[T_{GI},T\right]$ is
directly a segment of an arbitrary OFDM signal in (\ref{OFDM}), the whole
sampled discrete time sequence of $s(t),\ 0\leq t\leq T+T_{GI}$:
$s_n=s_n(nT_s),\ 0\leq n\leq N+M-2$, that is zero at the head and tail ends
from the above design idea of $s(t)$, may not be from a sampling of any OFDM
pulse in (\ref{OFDM}) for $t\in \left[0,T+T_{GI}\right]$. Thus, such a pulse
may not be used in the IRCI free range reconstruction as in \cite{TxzOFDMSAR}.
The key of this paper is to generate such a pulse $s(t)$ with the above
property of zero-valued head and tail, and in the meantime, its sampled
discrete time sequence $s_n$ is also a sampled discrete time sequence of an
OFDM pulse in (\ref{OFDM}) for $t\in \left[0,T+T_{GI}\right]$. Since the
non-zero pulse length is $T-T_{GI}$ and $T$ is arbitrary, the non-zero pulse
length is also arbitrary and independent of a swath width. The details is given
in the following subsections.

\subsection{Received signal model}
In order to better understand the IRCI free range reconstruction, let us first
see the receive signal model. Going back to (\ref{Receive}), for the $m$th
range cell, $R_m(\eta)=R_0(\eta)+m\rho_r$, where $R_0(\eta)$ is the
instantaneous slant range between the radar and the first range cell in the
swath as in \cite{TxzOFDMSAR}. Then, the part $t-\frac{2R_m(\eta)}{c}$ in
(\ref{Receive}) is equivalent to $t-\frac{2R_m(\eta)}{c}=t-t_0(\eta)-mT_s$,
where the constant time delay $t_0(\eta)=\frac{2R_0(\eta)}{c}$ is independent
of $m$ for a given slow time $\eta$. Let the sampling be aligned with the start
of the received signal after $t_0(\eta)$ seconds for the first arriving version
of the transmitted signal, $u(t,\eta)$ in (\ref{Receive}) can be converted to
the discrete time linear convolution of the transmitted sequence with the
weighting RCS coefficients $d_m$, i.e.,
\begin{equation}\label{uig}
  \tilde{u}_n=\sum_{m=0}^{M-1}d_ms_{n-m}+\tilde{w}_n,\ n=0,1,\ldots,N+2M-3,
\end{equation}
where
\begin{equation}\label{dm}
d_m=g_m\textrm{exp} \left\{-j4\pi
f_c\frac{R_m(\eta)}{c}\right\},\\
\end{equation}
in which $4\pi f_c\frac{R_m(\eta)}{c}$ in the exponential  is the azimuth
phase, and $s_n$ is the sampled discrete time sequence, $s_n=s(nT_s)$, of the
transmitted pulse $s(t)$ during $t\in \left[0,T+T_{GI}\right]$ for $T=NT_s$
and $T_{GI}=(M-1)T_s$. Since the range reconstruction in the SAR imaging
algorithm proposed in \cite{TxzOFDMSAR} in the following is only based on the
discrete time signal model in (\ref{uig}), what matters
in the range reconstruction is the discrete time sequence $s_n=s(nT_s)$, where
$s_n=0$ for $n<0$. If the sequence
${\bs'}=\left[s_0,s_{1},\ldots,s_{N+M-2}\right]^T$ in (\ref{uig}) has the
following zero head and tail property:
\begin{equation}\label{zeroHT}
\left[s_{0},\ldots,s_{M-2}\right]^T=\left[s_{N},\ldots,s_{N+M-2}\right]^T=
\mathbf{0}^{(M-1)\times 1},
\end{equation}
then, in terms of the range reconstruction later, the transmitted pulse $s(t)$
is equivalent to that with $s(t)=0$ for $t\in [0, T_{GI})$ and $t\in \left(T,
T+T_{GI}\right]$. It is also equivalent to an OFDM pulse in (\ref{OFDM}) such
that its sampled version
\begin{equation}\label{OFDMn}
s_n=s(nT_s)=\frac{1}{\sqrt{N}}\sum_{k=0}^{N-1}S_k
\textrm{exp}\left\{\frac{j2\pi kn}{N}\right\},\ n=0,1,\ldots,N+M-2,
\end{equation}
has the property (\ref{zeroHT}).

In summary, our proposed transmitted pulse of an arbitrary length $s(t)$ of
non-zero is that $s(t)=0$ for $t\in [0, T_{GI})$ and $t\in \left(T,
T+T_{GI}\right]$ and $s(t)$ has the OFDM form (\ref{OFDM}) for $t\in
\left[T_{GI}, T\right]$ with an arbitrary $T$ of $T>T_{GI}$, where the sampled
version $s_n$ of the analog waveform/pulse in (\ref{OFDM}) satisfies the zero
head and tail property (\ref{zeroHT}). Note that, since $T-T_{GI}$ is arbitrary
and $s(t)$ is only non-zero in the interval $\left[T_{GI},T\right]$, its
non-zero pulse length is arbitrary. Furthermore, since for the sequence $\bs'$,
its both head and tail parts are the same of all zeroes with length $M-1$, the
head part is a CP of the tail part and thus it fits to the sufficient CP based
SAR imaging proposed in \cite{TxzOFDMSAR}.

Based on the above analysis, in what follows, we  assume that an OFDM pulse in
(\ref{OFDM}) satisfies the zero head and tail property (\ref{zeroHT}) for its
sampled discrete time sequence $s_n$ and thus, it is equivalent to a pulse of
length $T-T_{GI}$ as described above in terms of the range reconstruction. So,
for convenience,  we may use these two kinds of pulses interchangeably. Note
that the reason why these two kinds of analog  waveforms are not the same is
because a non-zero OFDM signal in (\ref{OFDM}) can not be all zero for $t$ in
any interval of a non-zero length.

From (\ref{OFDMn}), it is clear that the time domain OFDM sequence
$\bs=\left[s_{0}, s_1,\ldots, s_{N-1}\right]^T$ is just  the $N$-point IFFT of
the vector $\bS=\left[S_0,S_1,\ldots,S_{N-1}\right]^T$. In the SAR imaging
algorithm proposed in \cite{TxzOFDMSAR}, $N\geq M$ is required, which is the
same as $T>T_{GI}$. However, from the above study, there are only $N-M+1$
non-zero values in the sequence ${\bs}$ and $N$ can be arbitrary as long as
$N\geq M$. In this case, the transmitted sequence is just
$\bs_t=\left[s_{M-1},s_{M},\cdots,s_{N-1}\right]^T\in \mathbb{C}^{(N-M+1)\times
1}$. Then, the first and the last $M-1$ samples of the received signal
$\tilde{u}_n$ in (\ref{uig}) do not contain any useful signal\footnote{In
\cite{TxzOFDMSAR}, the first and the last $M-1$ samples of the received signal
$\tilde{u}_n$ in (\ref{uig}) contain received target energy (or useful signal),
but they are redundant and removed at the receiver to obtain $u_n$ and IRCI
free range reconstruction.}, $d_m$. Thus, we can start the sampling at
$\tilde{u}_{M-1}$ as
\begin{equation}\label{un}
  u_n=\sum_{m=0}^{M-1}d_ms_{n-m+M-1}+w_n,\ n=0,1,\ldots,N-1.
\end{equation}
Now the question is how to design such an arbitrary length pulse, which is
studied next after the range reconstruction algorithm is introduced.

\subsection{Range compression}
In this subsection, we develop the range compression according to the above
OFDM received signal model. The received signal
$\bu=\left[u_{0},u_{1},\ldots,u_{N-1}\right]^T$ in (\ref{un}) is equivalent to
the following representation
\begin{equation}\label{resapeM}
  \bu=\bH\bs_t+\bw,
\end{equation} where $\bw=\left[w_{0},w_{1},\ldots, w_{N-1}\right]^T$ is the noise
vector and $\bH$ is the $N$ by $N-M+1$ matrix:
\begin{equation}\label{Hmatrix}
  \bH=
  \begin{bmatrix}
  d_0     & 0      & \cdots & 0      \\
  d_{1}   & d_0    & \ddots & \vdots \\
  \vdots  & \vdots & \ddots & 0      \\
  d_{M-1} & d_{M-2}& \cdots & \vdots \\
  0       & \ddots & \ddots & \vdots \\
  \vdots  & \ddots & d_{M-1}& d_{M-2}\\
  0       & \cdots & 0      & d_{M-1}
  \end{bmatrix}.
\end{equation}

The OFDM demodulator then performs the $N$-point  fast Fourier transform (FFT) on the vector
$\bu$:
\begin{equation}\label{U_i}\begin{array}{ll}
  U_i&=\frac{1}{\sqrt{N}}\sum\limits_{n=0}^{N-1}u_{n}\textrm{exp}
  \left\{\frac{-j2\pi in}{N}\right\}\\
  &=D_iS_i'+W_i,\ i=0,1,\ldots,N-1,
\end{array}\end{equation}
where $\left[S_0',S_1',\cdots,S_{N-1}'\right]^T$ is the $N$-point FFT of the
sequence $\left[\bs_t^T,\mathbf{0}^{1\times M-1}\right]^T$, a cyclic shift of
the time domain OFDM sequence $\bs$ of amount $M-1$, i.e.,
\begin{equation}\label{SiPrime}
S_i'=S_i\textrm{exp}\left\{\frac{j2\pi i(M-1)}{N}\right\},
\end{equation}
$\bW=\left[W_0,\ldots,W_{N-1}\right]^T$ is the $N$-point FFT of the noise
vector $\bw$, and
\begin{equation}\label{eDiM}
D_i=\sum_{m=0}^{M-1}d_{m}\textrm{exp}\left\{\frac{-j2\pi mi}{N}\right\}.
\end{equation}

Then, the estimate of $D_i$ is
\begin{equation}\label{hatD_i}
\hat{D}_i=\frac{U_i}{S_i'}=D_i+\frac{W_i}{S_i'},\ i=0,1,\ldots,N-1.
\end{equation}
The vector $\bD=\left[D_{0},D_1,\ldots,D_{N-1}\right]^T$ is just the $N$-point
FFT of the vector $\sqrt{N} \bgamma$, where $\bgamma$ is
\begin{equation}\label{dms}
\bgamma=\left[d_0,d_1,\cdots,d_{M-1},\underbrace{0,\cdots,0}_{N-M}\right]^T.
\end{equation}
So, the estimate of $d_m$ can be achieved by the $N$-point IFFT of the vector
$\hat{\bD}=\left[\hat{D}_{0},\hat{D}_1,\ldots,\hat{D}_{N-1}\right]^T$:
\begin{equation}\label{hatdm1}
  \hat{d}_{m}=\frac{1}{\sqrt{N}}\sum_{i=0}^{N-1}\hat{D}_i\textrm{exp}
  \left\{\frac{j2\pi mi}{N}\right\},\ m=0,\ldots, M-1.\\
\end{equation}
Then, we obtain the following estimates of the $M$ weighting RCS coefficients:
\begin{equation}\label{hatdm2}
    \hat{d}_m={\sqrt{N}}d_m+\hat{w}_m',\ m=0,\ldots, M-1,
\end{equation}
where $\hat{w}_m'$ is from the noise. In $(\ref{hatdm2})$, $d_m$ can be
 recovered without any IRCI from other range cells.

After the range compression, combining the equations (\ref{Receive})-(\ref{dm})
and (\ref{hatdm2}), we obtain
$$
\hat{g}_m= \hat{d}_m \textrm{exp} \left\{j4\pi
f_c\frac{R_m(\eta)}{c}\right\},
$$
and the range compressed signal can be written as
\begin{equation}\label{Range compression}
u_{ra}(t,\eta)=\sqrt{N}\sum_{m=0}^{M-1}\hat{g}_m
\delta\left(t-\frac{2R_m(\eta)}{c}\right)\textrm{exp}\left\{-j4\pi
f_c\frac{R_m(\eta)}{c}\right\}+w_{ra}(t,\eta),
\end{equation}
where $\delta\left(t-\frac{2R_m(\eta)}{c}\right)$ is the delta function with
non-zero value at $t=\frac{2R_m(\eta)}{c}$, which indicates that, for every
$m$, the estimate $\hat{g}_m$ of the RCS coefficient value $g_m$ is not
affected by any IRCI from other range cells after the range compression. In the
delta function, the target range migration is incorporated via the azimuth
varying parameter $\frac{2R_m(\eta)}{c}$. Also, the azimuth phase in the
exponential is unaffected by the range compression. In summary, the above range
compression provides an IRCI free range reconstruction.

Notice that unlike the processing in \cite{TxzOFDMSAR} where the first and the
last $M-1$ samples of the received signal are removed and thus cause
significant transmitted energy waste for a wide swath imaging, in the above
range reconstruction algorithm, all the transmitted energy is used for the
range compression without any waste. Since the transmitted OFDM pulse time
duration is $T-T_{GI}$, the minimum radar range is
$\frac{c\left(T-T_{GI}\right)}{2}$ that is also independent of a swath width.
Different from \cite{TxzOFDMSAR} where the CP part is not zero, the pulse
repetition interval $T_{PRI}$ becomes
$$
  T_{PRI}=\frac{1}{\textrm{PRF}}>\left(\frac{2R_w}{c}+\left(T-T_{GI}\right)\right),
$$
where $R_w$ is the swath width and PRF is the pulse repetition frequency (PRF).
We want to emphasize here that the minimum radar range and the maximum PRF of
our proposed OFDM SAR in this paper are the same as those in the existing SAR
systems, such as LFM SAR, when the same transmitted pulse time duration is used
\cite{Soumekh1999Synthetic, skolnik2001Introduction}.

In the above range compression, the processes of FFT in (\ref{U_i}), estimation
in (\ref{hatD_i}) and IFFT in (\ref{hatdm1}) are applied. Thus, it is necessary
to analyze the changes of the noise power in each step of the range
compression. Assume that $w_n$ in (\ref{un}) is a complex white Gaussian
variable with zero-mean and variance $\sigma^2$, i.e.,
$w_n\sim\mathcal{CN}\left(0,\sigma^2\right)$ for all $n$. Since the FFT
operation is unitary, the additive noise power does not change after the
process of (\ref{U_i}). Thus, $W_i$  also obeys
$W_i\sim\mathcal{CN}\left(0,\sigma^2\right)$ for all $i$. However, let
$\bar{W}_i=\frac{W_i}{S_i'}$ in (\ref{hatD_i}), then the variance of
$\bar{W}_i$ is changed to $\frac{\sigma^2}{\left|S_i\right|^2}$, where, from
(\ref{SiPrime}), $\left|S_i'\right|=\left|S_i\right|$, and thus
$\bar{W}_i\sim\mathcal{CN}\left(0,\frac{\sigma^2}{\left|S_i\right|^2}\right),\
i=0,\ldots,N-1$. Moreover, after the IFFT operation in (\ref{hatdm1}) we have
finished the range compression and the noise power of $\hat{w}_m'$ in
(\ref{hatdm2}) is
$\frac{\sigma^2}{N}\sum\limits_{i=0}^{N-1}\left|S_i\right|^{-2}$ and in the
meantime $\hat{w}_m'$, that follows the distribution
$\mathcal{CN}\left(0,\frac{\sigma^2}{N}\sum\limits_{i=0}^{N-1}
\left|S_i\right|^{-2}\right)$, is equivalent to the noise $w_{ra}(t,\eta)$ in
(\ref{Range compression}). Thus, from (\ref{hatdm2}), we can obtain the SNR of
the $m$th range cell after the range compression as,
\begin{equation}\label{SNRm}
\textrm{SNR}_m=\frac{N^2\left|d_m\right|^2}{\sigma^2\sum\limits_{i=0}^{N-1}
\left|S_i\right|^{-2}}.
\end{equation}
Notice that, we can obtain a larger $\textrm{SNR}_m$ with a smaller value of
$\sum\limits_{i=0}^{N-1}\left|S_i\right|^{-2}$ by designing $S_i$. With the
normalized energy constraint $\sum_{i=0}^{N-1}\left|S_i\right|^2=1$, when $S_i$
has constant module for all $i$, i.e.,
$\left|S_0\right|=\left|S_1\right|=\ldots=\left|S_{N-1}\right|=\frac{1}{\sqrt{N}}$,
we obtain the minimal value of
$\sum\limits_{i=0}^{N-1}\left|S_i\right|^{-2}=N^2$. In this case, the maximal
SNR after the range compression can be obtained as
\begin{equation}\label{SNRmax}
\textrm{SNR}_{max}=\frac{\left|d_m\right|^2}{\sigma^2}.
\end{equation}

Thus, the optimal signal
$S_i$ should have  constant module for all $i$, otherwise, the SNR after the
range
compression will be degraded. To evaluate the change of SNR, we define the SNR
degradation factor as
\begin{equation}\label{xi}
\xi=\frac{\textrm{SNR}_m}{\textrm{SNR}_{max}}
=\frac{N^2}{\sum\limits_{i=0}^{N-1}\left|S_i\right|^{-2}}.
\end{equation}
Notice that $\textrm{SNR}_m$ and $\textrm{SNR}_{max}$ are related to the $m$th
range cell in a swath, however, since $\xi\in\left(0,1\right]$ is independent
of the noise power $\sigma^2$ and $d_m$, the above  $\xi$ can be used to
evaluate the SNR degradation after the range compression for all range cells. A
larger $\xi$ denotes a less noise power enhancement (or a less SNR degradation)
caused by the estimation processing in (\ref{hatD_i}), and the generated signal
$S_i$ is closer to the optimal one.

Since the length of the transmitted OFDM sequence $\bs_t$ is $N_t=N-M+1$, from
the normalized energy constraint of $\bs_t$, the mean transmitted power of
$\bs_t$ is $\frac{1}{N_t}$. Thus, the SNR of the signal received from the $m$th
range cell before range reconstruction is
\begin{equation}
\overline{\textrm{SNR}}_{m}=\frac{\left|d_m\right|^2}{N_t\sigma^2}.
\end{equation}
We notice that the maximal SNR of the $m$th range cell after the range
compression $\textrm{SNR}_{max}$ in (\ref{SNRmax}) is equal to
$N_t\overline{\textrm{SNR}}_{m}$, and the range reconstruction SNR gain is the
same as that using LFM pulses with the same transmitted signal parameters
\cite{Soumekh1999Synthetic, skolnik2001Introduction}. However, because of the
sidelobes of the autocorrelation function using LFM pulses, the IRCI will occur
in the range reconstruction that degrades the signal-to-interference-plus-noise
ratio (SINR). Considering the $M$ range cells in a swath, the interference of
the $m$th range cell from other range cells in the swath is
\begin{equation}
  \textrm{I}_m=\sum_{
k=\textrm{max}\left\{-m,\ -({N_t}-1)\right\},\ k\neq 0
  }^{\textrm{min}\left\{M-m-1,\ {N_t}-1\right\}}
  d_{m+k}z(k),
\end{equation}
where $z(k)$ is the autocorrelation function of the LFM pulse, i.e.,
\begin{equation}
  z(k)=\sum_{n=0}^{{N_t}-1}l(n)l^*(n-k),\ \left|k\right|\leq {N_t-1},
\end{equation}
and $(\cdot)^*$ denotes the complex conjugate, $l(n),\ n=0,\ldots,{N_t}-1$, are
the values of a transmitted LFM sequence. ${N_t}$ denotes the length of the LFM
sequence that is equal to the length of the OFDM sequence we use in this paper.

Thus, the SINR of the signal after the range reconstruction using an LFM pulse
is
\begin{equation}
  \textrm{SINR}_m=\frac{\left|d_m\right|^2}{\left|\textrm{I}_m\right|^2+\sigma^2}.
\end{equation}

To investigate the mean SINR, for convenience, we consider the mean power of
range cells as $E\left[d_md_m^*\right]=\sigma_d^2$. Then, the mean interference
power, caused by the sidelobes, of each range cell in the swath is
\begin{equation}\label{EIm}
E\left[\left|\textrm{I}_m\right|^2\right]=\sigma_d^2\sum\limits_{
k=\textrm{max}\left\{-m,\ -({N_t}-1)\right\},\ k\neq 0
  }^{\textrm{min}\left\{M-m-1,\ {N_t}-1\right\}}
  \left|z(k)\right|^2.
\end{equation}

In this case, the mean SINR of the signal after the range reconstruction using
an LFM pulse is
\begin{equation}\label{SINRLFMmin}
  \textrm{SINR}_\textrm{LFM}=\frac{\sigma_d^2}
  {E\left[\left|\textrm{I}_m\right|^2\right]+\sigma^2}.
\end{equation}
For given $M$ and $N_t$, $\textrm{SINR}_\textrm{LFM}$ versus
$\frac{\sigma_d^2}{\sigma^2}$ can be calculated using
(\ref{EIm})-(\ref{SINRLFMmin}) and will be shown in the next section of
simulations. Notice that since a random sequence has the same level of the
sidelobe magnitudes of the autocorrelation values as an LFM signal does
\cite{TxzOFDMSAR}, the above SINR analysis also applies to the range
reconstruction in the  random noise SAR imaging.

In contrast, for the IRCI free range reconstruction by using an OFDM pulse, the
SINR is equal to the SNR of the signal after the range reconstruction, since
for every range cell, there is no inter-range-interference from other range
cells. If the lower bound of the module of the OFDM sequence $\bS$ is
$S_{min}$, i.e., $\left|S_i\right|> S_{min}$ for all $i=0,1,\ldots,N-1$, we can
obtain
$$
\sum\limits_{i=0}^{N-1}\left|S_i\right|^{-2}<NS_{min}^{-2}.
$$
Then, from (\ref{SNRm}), the SNR for the $m$th range cell signal is lower
bounded by
\begin{equation}\label{SNROFDMmin}
\textrm{SNR}_m=\frac{N^2\left|d_m\right|^2}
{\sigma^2\sum\limits_{i=0}^{N-1}\left|S_i\right|^{-2}}>
\frac{N\left|d_m\right|^2}{\sigma^2S_{min}^{-2}}.
\end{equation}
Thus, the SINR for all range cells after the range reconstruction is also lower
bounded by
\begin{equation}\label{SINROFDMmin}
\textrm{SINR}_\textrm{OFDM}=E\left[\textrm{SNR}_m\right]=\frac{N^2\sigma_d^2}
{\sigma^2\sum\limits_{i=0}^{N-1}\left|S_i\right|^{-2}}>
\frac{N\sigma_d^2}{\sigma^2S_{min}^{-2}}.
\end{equation}
A remark to the lower bound for the SINR in (\ref{SINROFDMmin}) is that it does
not depend on the swath width $M$, which is because our proposed OFDM SAR
imaging algorithm with our proposed arbitrary length OFDM pulses is IRCI free
and the pulse length does not depend on a swath width. Therefore, it is
particularly interesting in wide swath SAR imaging applications. Based on the
above analysis, the task here is to generate an OFDM sequence with a larger
$\xi$ (or a less SNR degradation) by designing a sequence $S_i$ with larger
$S_{min}$. This motivates the following OFDM sequence design.


\section{New OFDM Sequence Design}\label{sec4}

First of all, an OFDM pulse of any segment in (\ref{OFDM}) is determined by a
weight sequence ${\bS}=\left[S_0, S_1,\ldots,S_{N-1}\right]^T$ that is
determined by its $N$-point IFFT ${\bs}=\left[s_0,s_1,\ldots,s_{N-1}\right]^T$.
Thus, an OFDM pulse design is equivalent to the design of its weight sequence
${\bS}$ or the $N$-point IFFT, ${\bs}$, of ${\bS}$. From the studies in the
preceding sections, an arbitrary length OFDM pulse $s(t)$ supported only in
$\left[T_{GI}, T\right]$ for $T>T_{GI}$ with its sampled sequence $s_n=s(nT_s)$
should be designed as follows.

1) Sequence ${\bs}$ should satisfy the zero head condition in (\ref{zeroHT}).
When this condition is satisfied and the $N$-point FFT, ${\bS}$, of ${\bs}$, is
used as the weight sequence in the OFDM pulse in (\ref{OFDM}) denoted as
$s_1(t)$, let its segment (or truncated version) only supported on
$\left[T_{GI}, T\right]$ be denoted by $s(t)$ that is $0$ for $t\in \left[0,
T_{GI}\right)\cup \left(T, T+T_{GI}\right]$ and equals $s_1(t)$ for $t\in
\left[T_{GI}, T\right]$. Then, pulse $s(t)$ is still an OFDM pulse on its
support and has length $T-T_{GI}$ of support (i.e., non-zero values) and this
length can be arbitrary and independent of a swath width. Furthermore, $s(t)$
has the same discrete-time sequence ${\bs'}$ as the OFDM pulse $s_1(t)$ does,
which, thus, satisfies the zero head and tail condition (\ref{zeroHT}). From
the study in the preceding section, transmitting pulse $s(t)$ leads to the IRCI
free range reconstruction in SAR imaging.

2) To avoid enhancing the noise as the estimation processing in (\ref{hatD_i})
and achieve the maximal possible SNR after the range compression, the complex
weights $S_i$ should be as constant module as possible for all $i$. In other
words, $S_{min}$ should be as large as possible.

3) The PAPR of the transmitted OFDM pulse $s(t)$ in (\ref{OFDM}) for $t\in
\left[T_{GI},T\right]$ should be minimized so that its transmitting and
receiving can be implemented easier. Otherwise, a delta pulse would serve 1)
and 2) above, but it has infinite bandwidth and infinite PAPR and can not be
transmitted \cite{skolnik2001Introduction}.

Unfortunately, it looks like that there is no closed-form solution of an OFDM
sequence ${\bs}$ that simultaneously satisfies the above requirements 1)-3). It
would be easy to have a sequence ${\bs} =\left[s_0,s_1,\ldots,s_{N-1}\right]^T$
to satisfy the zero head condition in (\ref{zeroHT}), i.e., $s_n=0$ for
$n=0,1,\ldots,M-2$ as mentioned in the above 1). However, its FFT, ${\bS}$, may
not have constant module or may not be even close to constant module. A natural
idea is to modify this sequence ${\bS}$ to be closer to constant module and
then take its IFFT to go back to the time domain ${\bs}$ and also obtain the
continuous waveform $s(t)$. Then, this ${\bs}$ may not satisfy the zero head
condition in (\ref{zeroHT}) anymore. Furthermore, the PAPR of the continuous
waveform $s(t)$ for $t \in \left[T_{GI}, T\right]$ may be high. In this case,
we may modify ${\bs}$ and in the meantime add some constraint to limit the PAPR
of $s(t)$ for $t \in \left[T_{GI}, T\right]$. Our OFDM sequence design idea is
to do the above process iteratively until a pre-set iteration number is reached
and/or a desired sequence ${\bs}$ is obtained.

\begin{figure}[t]
\begin{center}
\includegraphics[width=0.45\columnwidth,draft=false]{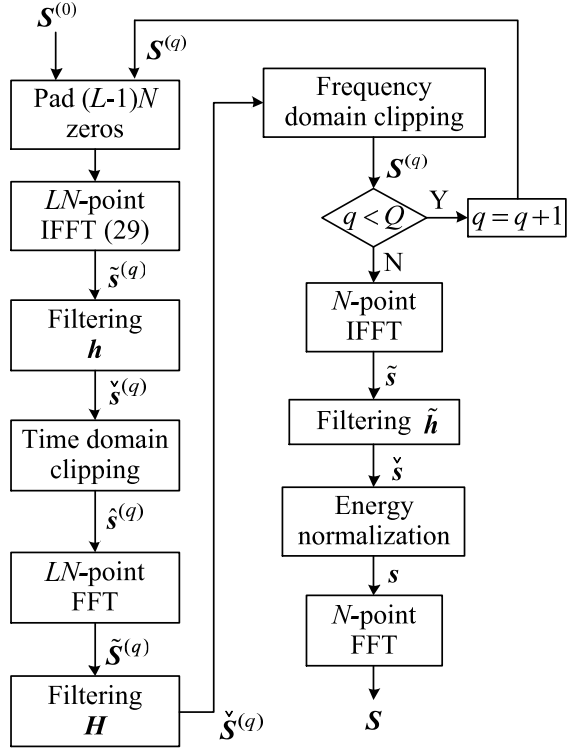}
\end{center}
\caption{Block diagram of the OFDM sequence design algorithm.}\label{ICF_block}
\end{figure}

To clearly describe the design algorithm, let us better understand the PAPR
calculation for an analog waveform. For a sufficiently accurate PAPR estimation
of a transmitted OFDM pulse, we usually consider its oversampled discrete time
sequence, i.e., a time domain OFDM sequence
$\tilde{\bs}=\left[\tilde{s}_0,\ldots,\tilde{s}_{LN-1}\right]^T$ by $L$ times
over-sampling of the continuous waveform $s(t)$ with complex weights
$\bS=\left[S_0,S_1,\ldots,S_{N-1}\right]^T$ in (\ref{OFDM}) for a sufficiently
large $L$ \cite{SeungIEEEWCPAPRoverview200556}, i.e.,
\begin{equation}\label{overSamplingT}
 \tilde{s}_n=\frac{1}{\sqrt{LN}}\sum_{i=0}^{N-1}S_i\textrm{exp}
 \left\{\frac{j2\pi ni}{LN}\right\},\ n=0,\ldots, LN-1,
\end{equation}
which can be implemented by the $LN$-point IFFT of the sequence $\left[S_0,
S_1,\ldots,S_{N-1}, 0, 0,\ldots, 0\right]^T$ of length $LN$.

Then, the PAPR of the transmitted OFDM pulse can be defined as
\begin{equation}\label{PAPR}
  \textrm{PAPR}=\frac{\underset{n=0,\ldots,LN-1}{\mathop{\textrm{max}}}
  \left|\tilde{s}_n\right|^2}{\frac{1}{LN}\sum_{n=0}^{LN-1}\left|\tilde{s}_n\right|^2}.
\end{equation}

Since ${\bs}$ and ${\bS}$ are FFT pairs, starting with ${\bs}$ and starting
with ${\bS}$ are equivalent. For the convenience to deal with the PAPR issue,
our proposed iterative algorithm starts with an initial random constant modular
sequence ${\bS}^{(0)}\in\mathbb{C}^{N\times 1}$ and obtains
$\tilde{\bs}^{(q)}\in\mathbb{C}^{LN\times 1}$ using (\ref{overSamplingT}) as
shown in Fig. \ref{ICF_block}.

Since the first $M-1$ samples of our desired sequence $\bs$ should be equal to
zero, after the $L$ times over-sampling of the analog waveform, the first
$L(M-1)$ samples in sequence $\tilde{s}_n^{(q)},\ 0\leq n\leq L(M-1)-1$, should
be equal to zero. Thus, we apply the following time domain filter to the newly
obtained sequence $\tilde{\bs}^{(q)}$:
\begin{equation}
  h(n)=\left\{\begin{array}{ll}
 0,\ 0\leq n\leq L(M-1)-1\\
 1,\ L(M-1)\leq n\leq LN-1
  \end{array}\right.,
\end{equation}
as $\check{s}^{(q)}_n=\tilde{s}_n^{(q)}h(n),\ n=0,\ldots,LN-1$, to obtain a new
sequence
$\check{\bs}^{(q)}=\left[\check{s}^{(q)}_0,\ldots,\check{s}^{(q)}_{LN-1}\right]^T$.
After this truncation, we then add a PAPR constraint to the segment of the
non-zero elements of this sequence to obtain the next new sequence
$\hat{s}_n^{(q)}$ by clipping $\check{s}_n^{(q)}$ as follows. The time domain
clipping can be defined as, \cite{ArmstrongELPAPRreduction2002246},
\begin{subequations}
\begin{align}
\hat{s}^{(q)}_n&=\left\{\begin{array}{ll}
\textrm{T}_q\frac{\check{s}^{(q)}_n}{|\check{s}^{(q)}_n|},&\textrm{if}\
|\check{s}^{(q)}_n|>\textrm{T}_q\\
\check{s}^{(q)}_n,&\textrm{if}\ |\check{s}^{(q)}_n|\leq \textrm{T}_q
\end{array}\right.,\\
\textrm{T}_q&=\sqrt{\textrm{PAPR}_dP_{tav}^{(q)}},
\end{align}\end{subequations}
where $L(M-1)\leq n\leq LN-1$, and $\hat{s}^{(q)}_n=0$ for
$n=0,\ldots,L(M-1)-1$.
\[P_{tav}^{(q)}=\frac{1}{L(N-M+1)}{\sum\limits_{n=L(M-1)}^{LN-1}
\left|\check{s}^{(q)}_n\right|^2}\] is the average power of the non-zero
elements in sequence $\check{\bs}^{(q)}$. $\textrm{T}_q$ is the clipping level
in the $q$th iteration which is updated in each iteration according to the
average power $P_{tav}^{(q)}$ and a constant value $\textrm{PAPR}_d$ that is a
lower bound for a desired PAPR.

After the $LN$-point FFT operation to $\hat{s}^{(q)}_n$, we obtain the
frequency domain sequence $\tilde{\bS}^{(q)}$. To constrain the out-of-band
radiation caused by the time domain filtering and clipping, we also use a
filter in the frequency domain:
\begin{equation}
  H(i)=\left\{\begin{array}{ll}
 1,\ 0\leq i\leq N-1\\
 0,\ N\leq i\leq LN-1
  \end{array}\right..
\end{equation}
And the output sequence $\check{S}_i^{(q)}$ can be obtained by
$\check{S}^{(q)}_i=\tilde{S}^{(q)}_iH(i),\ i=0,\ldots,LN-1$. To deal with the
constant module issue of the frequency domain sequence ${\bS}$, then, the
following frequency domain clipping is used:
\begin{equation}
{S}^{(q+1)}_i=\left\{\begin{array}{ll}
\sqrt{P_{fav}^{(q)}}\left(1+G_f\right)\frac{\check{S}^{(q)}_i}{|\check{S}^{(q)}_i|},
&\textrm{if}\
|\check{S}^{(q)}_i|>\sqrt{P_{fav}^{(q)}}\left(1+G_f\right)\\
\sqrt{P_{fav}^{(q)}}\left(1-G_f\right)\frac{\check{S}^{(q)}_i}{|\check{S}^{(q)}_i|},
&\textrm{if}\
|\check{S}^{(q)}_i|<\sqrt{P_{fav}^{(q)}}\left(1-G_f\right)\\
\check{S}^{(q)}_i,&\ \textrm{otherwise}
\end{array}\right..
\end{equation}
where $0\leq i\leq N-1$, and sequence ${\bS}^{(q+1)}=\left[S^{(q+1)}_0,
S^{(q+1)}_1,\ldots, S^{(q+1)}_{N-1}\right]^T$ is obtained. And
\[P_{fav}^{(q)}=\frac{1}{N}\sum\limits_{i=0}^{N-1}\left|\check{S}^{(q)}_i\right|^2\] is the
average power of the non-zero elements in sequence $\check{\bS}^{(q)}$. $G_f$
is a factor that we use to control the upper and lower bounds for sequence
${S}^{(q+1)}_i$. Thus, the module of sequence ${S}^{(q+1)}_i$ is constrained as
$\left|{S}^{(q+1)}_i\right|\in \left[\sqrt{P_{fav}^{(q)}}\left(1-G_f\right),
\sqrt{P_{fav}^{(q)}}\left(1+G_f\right)\right]$. A smaller $G_f$ denotes that a
closer-to-constant modular sequence ${\bS}^{(q+1)}$ can be obtained.

The above procedure is done for $q=0, 1,\ldots$, when $q<Q$, where $Q$ is a
pre-set maximum iteration number. When $q=Q$, the iteration stops and then
$N$-point IFFT is applied to ${\bS}^{(Q)}\in\mathbb{C}^{N\times 1}$ to obtain
$\tilde{\bs}\in\mathbb{C}^{N\times 1}$. After that, a time domain filter, i.e.,
\[\tilde{h}(n)=\left\{\begin{array}{ll}
 0,\ 0\leq n\leq M-2\\
 1,\ M-1\leq n\leq N-1
\end{array}\right.,\]
is applied to $\tilde{\bs}$ to obtain sequence
$\check{\bs}=\left[\check{s}_0,\ldots,\check{s}_{N-1}\right]^T$, where
$\check{s}_{n}=\tilde{s}_n\tilde{h}(n),\ n=0,\ldots,N-1$. In order to normalize
the energy of the sequence ${\bs}$ to $1$, we use the normalization to the time
domain sequence $\check{\bs}$ as
\[s_n=\frac{\check{s}_n}{\sqrt{\sum\limits_{k=M-1}^{N-1} \left|\check{s}_{k}\right|^2}},\
n=0,\ldots,N-1,\] and obtain the OFDM sequence $\bs$ in (\ref{OFDMn}) that
satisfies the zero head condition in (\ref{zeroHT}). Finally, $\bS$ can be
obtained by taking the $N$-point FFT of $\bs$. The PAPR of the non-zero part of
$s_n$ for $M-1\leq n\leq N-1$ can be calculated using (\ref{overSamplingT}) and
(\ref{PAPR}) and the noise power enhancement factor $\xi$ in (\ref{xi}) can
also be calculated from $\bS$.

Notice that, after the last iteration, the filtering operation in time domain
is applied to $\tilde{\bs}$ to obtain ${\bs}$, which will cause some
out-of-band radiation to $\bS$. However, comparing to the OFDM sequence energy,
the out-of-band radiation energy is much smaller and can be ignored as we shall
see later in the simulations in the next section.

Therefore, for a given swath width and radar range resolution, we can obtain
$M$. Then, for any $N$ with $N\geq M$, by using the above OFDM pulse design
method, we can obtain an OFDM sequence $\bs$ with $M-1$ zeros at the head part
of $\bs$ and $N-M+1$ non-zero values in the remaining part of $\bs$, and also
its $N$-point FFT $\bS$. With this $\bS$ as the weights in (\ref{OFDM}), the
OFDM pulse $s(t)$ in (\ref{OFDM}) for $t\in \left[T_{GI}, T\right]$ can be
obtained. Since $N$ or correspondingly $T$ can be chosen arbitrarily, the pulse
length, $T-T_{GI}$, of $s(t)$ can be arbitrary and independent of $M$ (or the
swath width).

Let us go back to the mean SINR in (\ref{SINROFDMmin}) using OFDM pulses. Note
that the constant module sequence ${S_i}$ is achieved when
$\left|S_i\right|=\frac{1}{\sqrt{N}}$ for all $i,\ i=0,1,\ldots,N-1$. According
to our numerous simulations, we find that it is not difficult to generate an
OFDM sequence $S_i$ with $\left|S_i\right|\geq 0.8\frac{1}{\sqrt{N}},\
i=0,\ldots,N-1$, using our proposed OFDM pulse design algorithm above, which
can be seen in the next section. Simulations about the above SINR comparison
are also provided in the next section.



\section{Simulation Results}\label{Simulation}
In this section, by using simulations we first see the performance of our
proposed OFDM sequence/pulse design of arbitrary length. We then see the
performance of the IRCI free range reconstruction in SAR imaging with our
proposed arbitrary length OFDM pulse.
\subsection{Performance of the OFDM pulse design}

\begin{figure}[b]%
\centering \subfigure[]{
\label{PAPR_Qs} 
\includegraphics[width=3.1in]{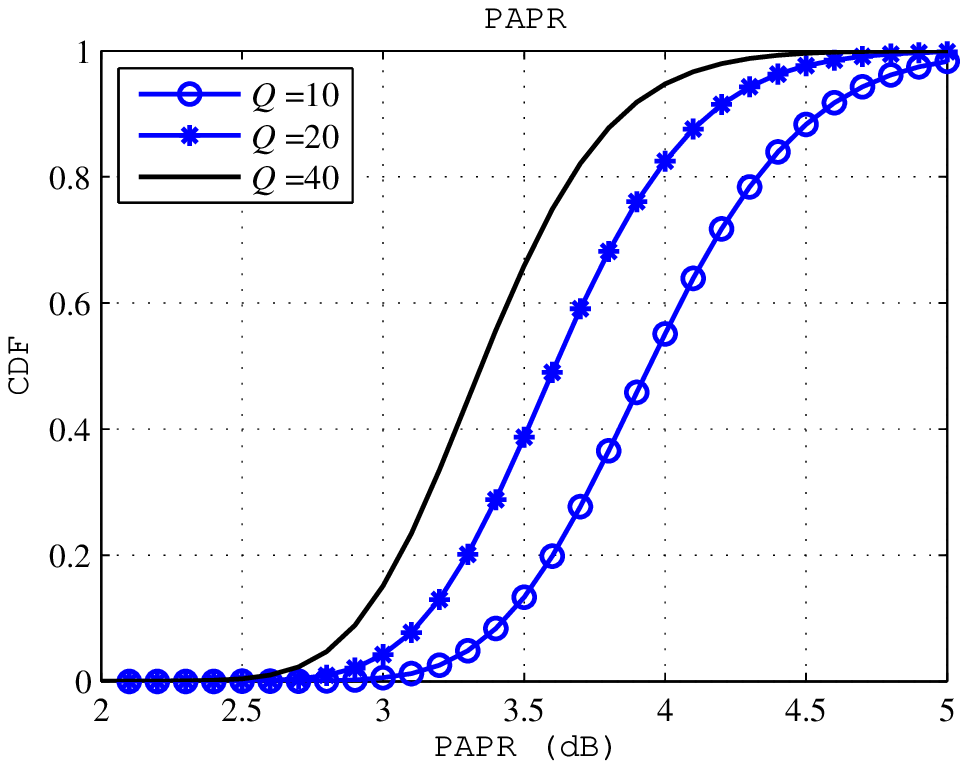}}
\hspace{0.02in} \subfigure[]{ \label{SNR_Qs} 
\includegraphics[width=3.1in]{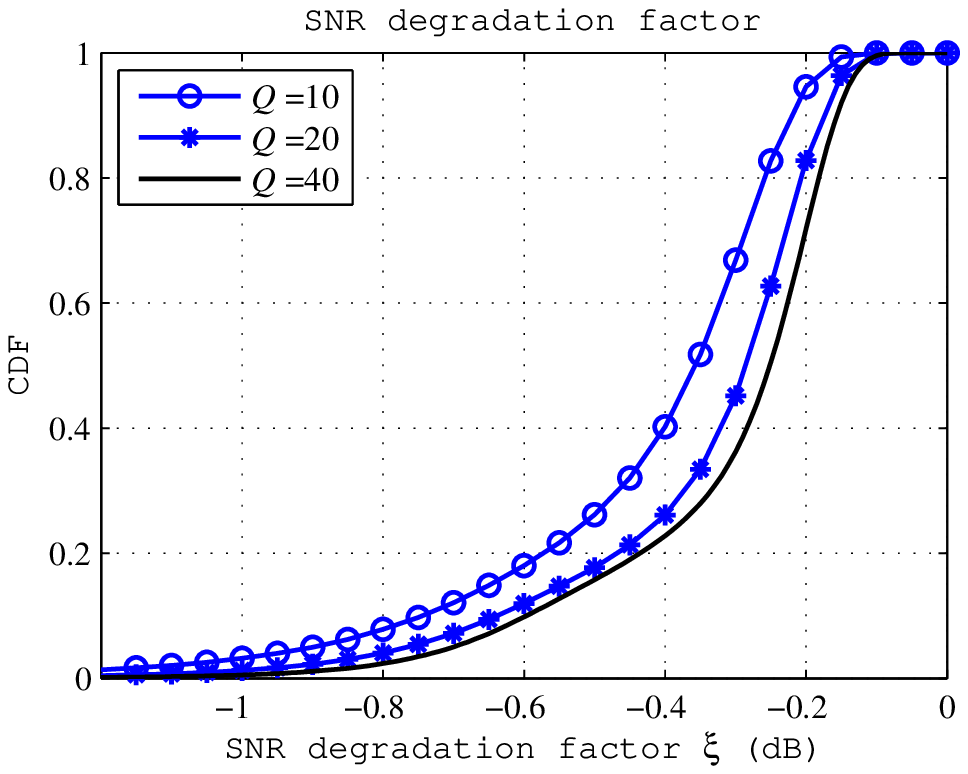}}
\caption{The CDFs for different $Q$ with $\textrm{PAPR}_d$ $=1$ dB and
$G_f=5\%$: (a) PAPR; (b) SNR degradation factor.}
\label{Qs} 
\end{figure}

\begin{figure}[t]%
\centering \subfigure[]{
\label{PAPR_PAPRs} 
\includegraphics[width=3.1in]{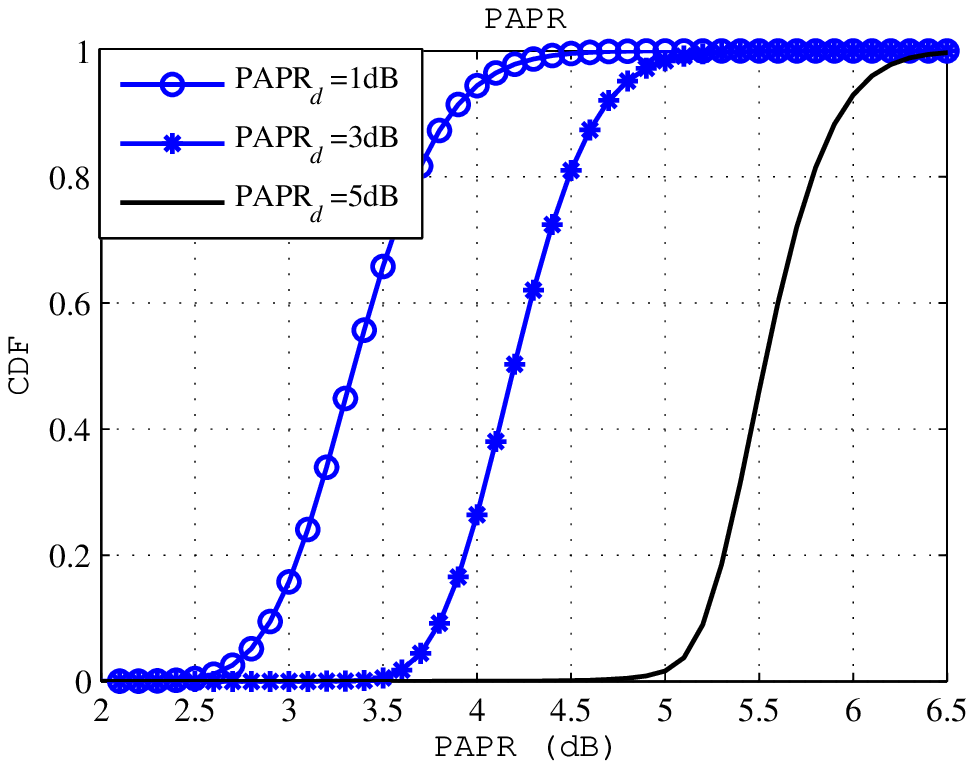}}
\hspace{0.02in} \subfigure[]{ \label{SNR_PAPRs} 
\includegraphics[width=3.1in]{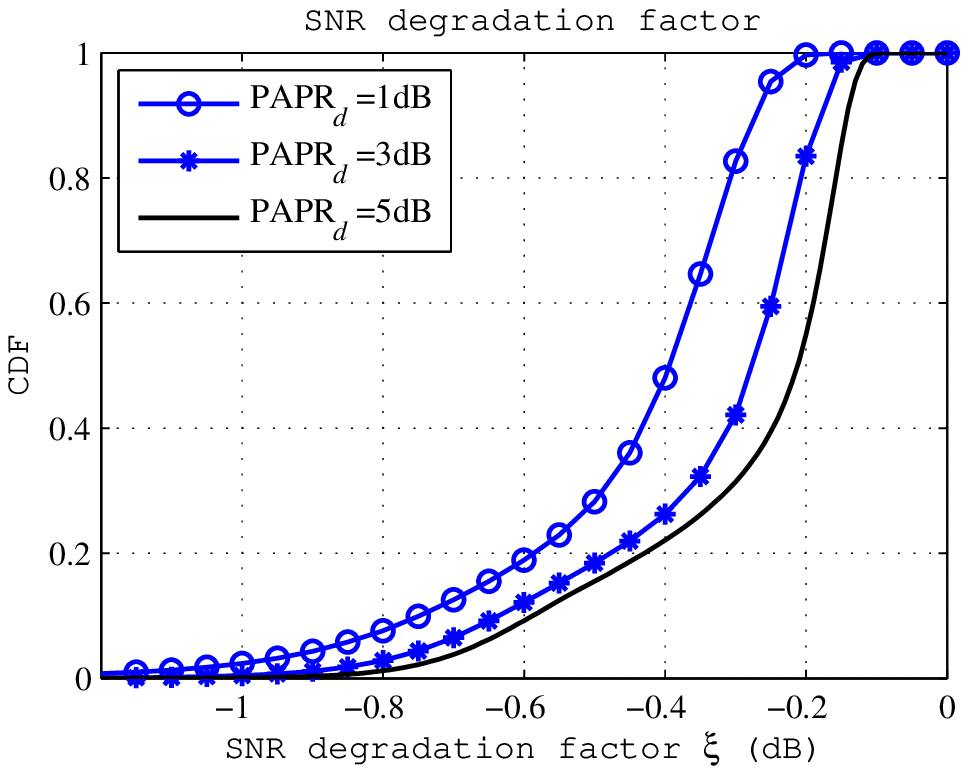}}
\caption{The CDFs for different $\textrm{PAPR}_d$ with $Q=20$ and $G_f=10\%$:
(a) PAPR; (b) SNR degradation factor.}
\label{PAPRs} 
\end{figure}

\begin{figure}[b]%
\centering \subfigure[]{
\label{PAPR_Gfs} 
\includegraphics[width=3.1in]{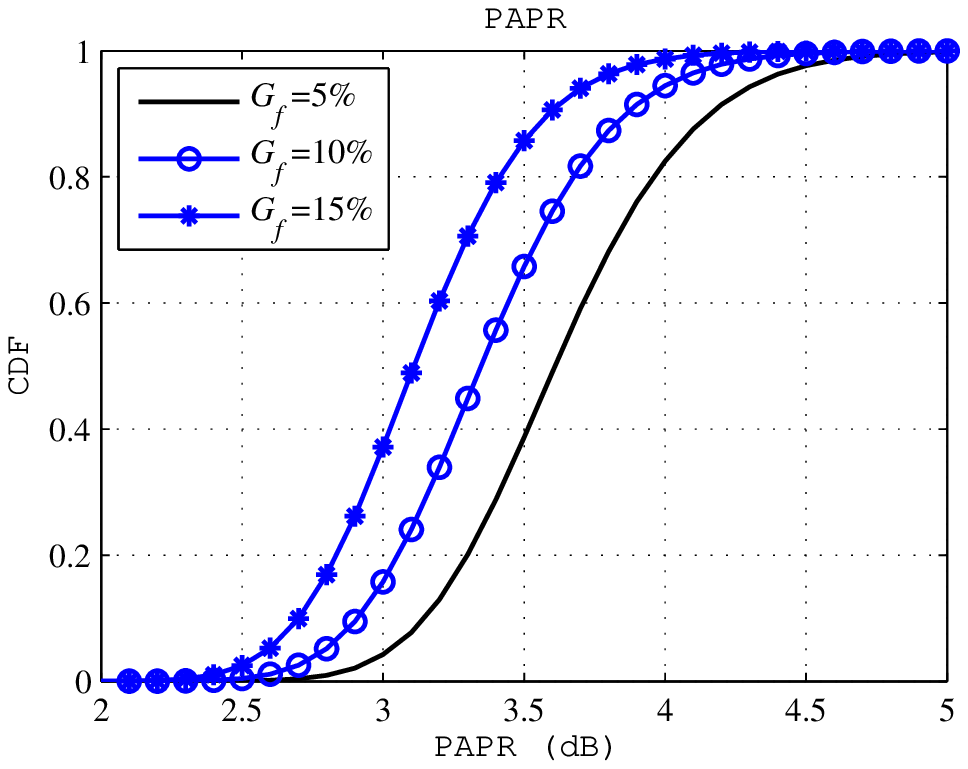}}
\hspace{0.02in} \subfigure[]{ \label{SNR_Gfs} 
\includegraphics[width=3.1in]{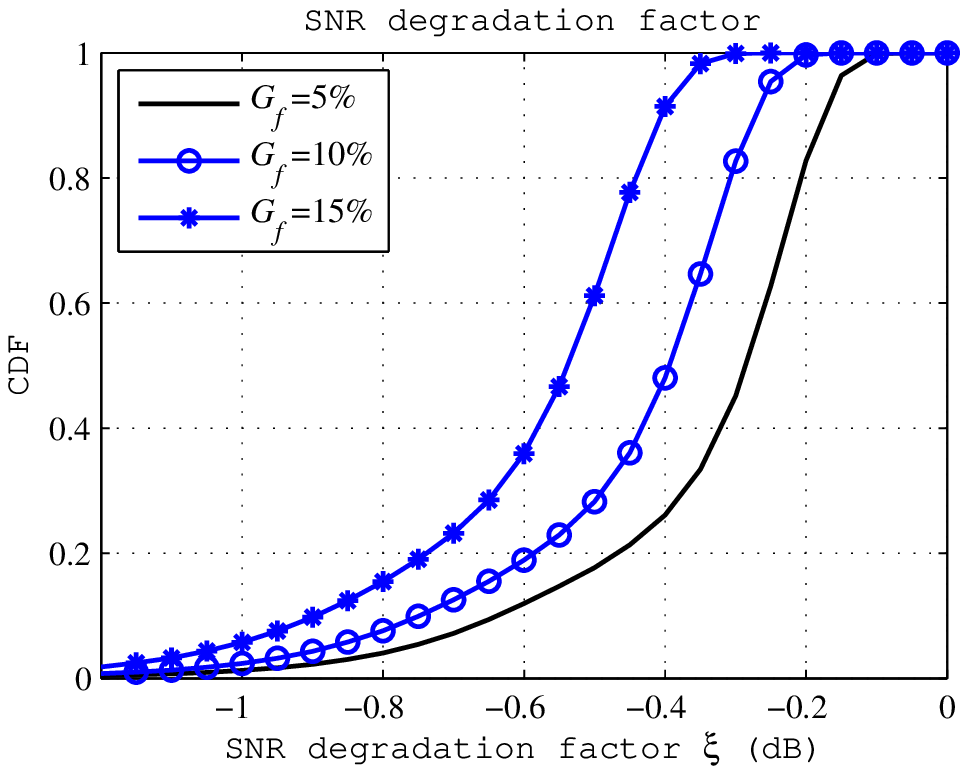}}
\caption{The CDFs for different $G_f$ with $Q=20$ and $\textrm{PAPR}_d$ $=1$
dB: (a) PAPR; (b) SNR degradation factor.}
\label{Gfs} 
\end{figure}

In this subsection, we first discuss the performance of the OFDM pulse design
algorithm. For simplicity, we set $M=96$ and $N=128$. To achieve a sufficiently
accurate PAPR estimate, we set the over-sampling ratio $L=4$
\cite{SeungIEEEWCPAPRoverview200556}. Then, we can generate an OFDM sequence
$\bs$ with $M-1=95$ zeros at the head part of $\bs$. We evaluate the PAPR and
the SNR degradation factor $\xi$ by using the standard Monte Carlo technique
with $5\times 10^5$ independent trials. In each trial, the $i$th element of
initial sequence ${\bS}^{(0)}$ is set as $S_i^{(0)}=e^{j2\pi \varphi_i},\
i=0,\ldots,N-1$, where $\varphi_i$ is uniformly distributed over the interval
$[0,2\pi]$. In Figs. \ref{Qs}-\ref{Gfs}, we plot the cumulative distribution
functions (CDF) of the PAPR and the SNR degradation factor $\xi$. The curves in
Fig. \ref{Qs} denote that, with the increase of the maximum iteration number
$Q$, the PAPR decreases and the $\xi$ increases to $1$. In Fig. \ref{PAPR_Qs},
more than $10\%$, $40\%$ and $60\%$ of the PAPRs of the OFDM sequences are less
than $3.5$ dB when $Q$ is equal to $10$, $20$ and $40$, respectively. In Fig.
\ref{SNR_Qs}, the probability of $\xi>-0.4\ \textrm{dB}\approx 0.91$, i.e.,
$\textrm{Pr}\left(\xi>0.91\right)=1-\textrm{Pr}\left(\xi\leq 0.91\right)$, is
about $60\%$, $75\%$ and $78\%$ for $Q$ is equal to $10$, $20$ and $40$,
respectively. $\xi>-0.4\ \textrm{dB}\approx 0.91$ denotes that the SNR of the
received signal after the range reconstruction (using the designed OFDM pulse)
is more than $91\%$ of the maximum SNR using constant modular weights $S_i$.
Thus, the SNR degradation of the CP based SAR imaging algorithm can be
insignificant by using our designed arbitrary length OFDM pulses. We also plot
the CDFs for different $\textrm{PAPR}_d$ with $Q=20$ and $G_f=10\%$ in Fig.
\ref{PAPRs}. The curves in Fig. \ref{PAPRs} show that the PAPR change is more
sensitive than the $\xi$ change for different $\textrm{PAPR}_d$. Specifically,
the curves in Fig. \ref{PAPR_PAPRs} indicate that the PAPR of a designed $\bs$
is significantly increased for the increase of $\textrm{PAPR}_d$. And the
curves in Fig. \ref{SNR_PAPRs} denote that the SNR degradation becomes less
when $\textrm{PAPR}_d$ is higher. Similarly, the curves in Fig. \ref{Gfs}
indicate that the PAPR of $\bs$ is decreased and the SNR degradation is
increased, when $G_f$ is increased.

\begin{table}[!htp]
\caption{The numbers of Monte Carlo trials for $\xi$ and PAPR with
$\textrm{PAPR}_d$ $=1$ \textnormal{dB} and $G_f=5\%$} \label{RunNumbers}
\begin{center}
\begin{tabular}{|c|c|c|c|}
\hline 
 &$\xi\geq -0.1$ dB  & $\xi\geq -0.2$ dB  & $\xi\geq -0.4$ dB \\
\cline{1-4}
PAPR $\leq 2$ dB  & 4   & 5     & 7  \\
\cline{1-4}
PAPR $\leq 2.5$ dB & 145 & 1511  & 2134 \\
\cline{1-4}
PAPR $\leq 3$ dB  & 615 & 35036 & 69735 \\
\hline
 \multicolumn{4}{|c|}{Total number of trails: $5\times 10^5$} \\
\hline
\end{tabular}
\end{center}
\end{table}

In practice, we want to generate an OFDM sequence $\bs$ with the minimal PAPR
as well as the minimal SNR degradation. However, according to the above
analysis the PAPR and $\xi$ are interacting each other. Therefore, it is
necessary to consider the constraints of both  PAPR and $\xi$ at the same time.
In Table \ref{RunNumbers}, we count the numbers of trails under different
conditions of the PAPR and $\xi$ within the $5\times 10^5$ Monte Carlo
independent trails for $Q=40$, $\textrm{PAPR}_d$ $=1$ dB and $G_f=5\%$.
Although only $4$ trails meet the constraints of PAPR$\leq 2$ dB and $\xi\geq
-0.1$ dB, it can also indicate that an OFDM sequence with both low PAPR and low
SNR degradation can be achieved by using our proposed OFDM pulse design
algorithm. We also count the numbers of trails under different conditions of
$S_{min}$ in Table \ref{SminNumbers}. The number of trails for $S_{min}\geq
0.8\frac{1}{\sqrt{N}}$ are $14415$, especially, there are $7$ trails with
$S_{min}\geq 0.88\frac{1}{\sqrt{N}}$. These results indicate that it is not
difficult to generate an OFDM sequence $\bS$ with $S_{min}\geq
0.8\frac{1}{\sqrt{N}}$. Specifically, a more excellent OFDM sequence with lower
PAPR, larger $\xi$, and larger $S_{min}$ can be obtained by doing more Monte
Carlo trails or with a larger iteration number $Q$, since in practice, the same
OFDM pulse is used for SAR imaging and can be generated off-line. In all of the
above simulations, the out-of-band radiation energy of $\bS$ is less than
$10^{-30}$ and thus it can be completely ignored.

\begin{table}[!htp]
\caption{The numbers of Monte Carlo trials for $S_{min}$ with $\textrm{PAPR}_d$
$=1$ \textnormal{dB} and $G_f=5\%$} \label{SminNumbers}
\begin{center}
\begin{tabular}{|c|c|c|c|}
\hline  %
$S_{min}\geq 0.88\frac{1}{\sqrt{N}}$  & $S_{min}\geq 0.85\frac{1}{\sqrt{N}}$
& $S_{min}\geq 0.8\frac{1}{\sqrt{N}}$ & $S_{min}\geq 0.5\frac{1}{\sqrt{N}}$\\
\hline
 7   & 371     & 14415 & 353782  \\
\hline
 \multicolumn{4}{|c|}{Total number of trails: $5\times 10^5$} \\
\hline
\end{tabular}
\end{center}
\end{table}

\begin{figure}[b]%
\centering \subfigure[]{
\label{SINR_mfig} 
\includegraphics[width=3.1in]{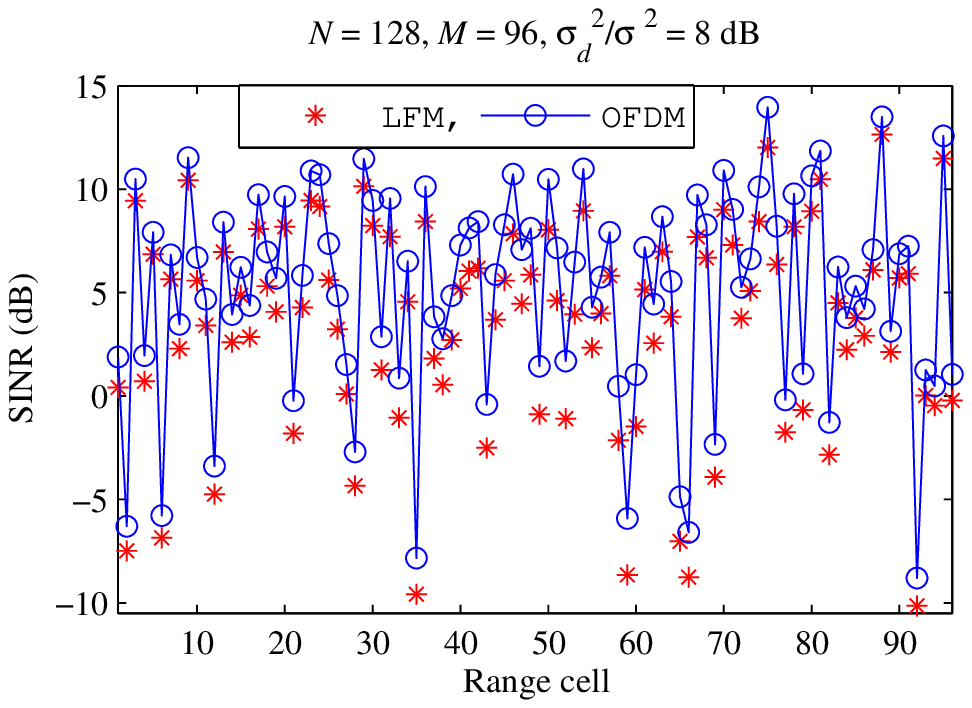}}
\hspace{0.02in} \subfigure[]{ \label{SINR_m_zoonfig} 
\includegraphics[width=3.1in]{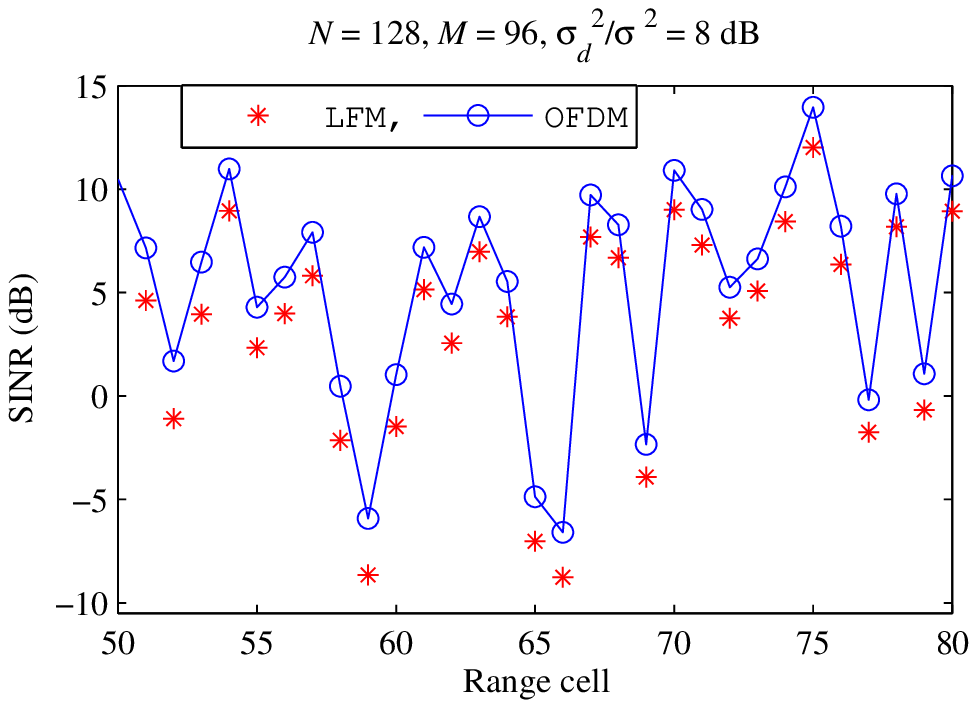}}
\caption{The SINRs after the range reconstructions using an LFM pulse and a
designed OFDM pulse: (a) SINRs of all the $M$ range cells; (b) The zoom-in
image of (a).}
\label{SINRm_LFM_OFDM} 
\end{figure}

We also investigate the SINRs of the signals after the range reconstructions by
using an LFM pulse and a designed OFDM pulse with $N=128$ in Fig.
\ref{SINRm_LFM_OFDM}. The parameters of the LFM pulse are the same as the OFDM
pulse, such as the transmitted pulse time duration, bandwidth and transmitted
signal energy. We randomly choose a designed OFDM sequence with PAPR $=1.84$
dB, $\xi=-0.11$ dB and $S_{min}=0.8\frac{1}{\sqrt{N}}$. The randomly generated
weighting RCS coefficients, $d_m,\ m=0,\ldots,M-1$, are included in $M=96$
range cells in a swath with $\frac{\sigma_d^2}{\sigma^2}=8$ dB. Then, the
transmitted sequence length is $N_t=33$ that is independent of $M$. The SINRs
of all the $M$ range cells are shown in Fig. \ref{SINR_mfig}. This figure
indicates that the SINRs by using a designed OFDM pulse are larger than the
SINRs by using an LFM pulse. The details from the $50$th range cell to the
$80$th range cell are shown in its zoom-in image in Fig. \ref{SINR_m_zoonfig}.

\begin{figure}[t]
\begin{center}
\includegraphics[width=0.65\columnwidth,draft=false]{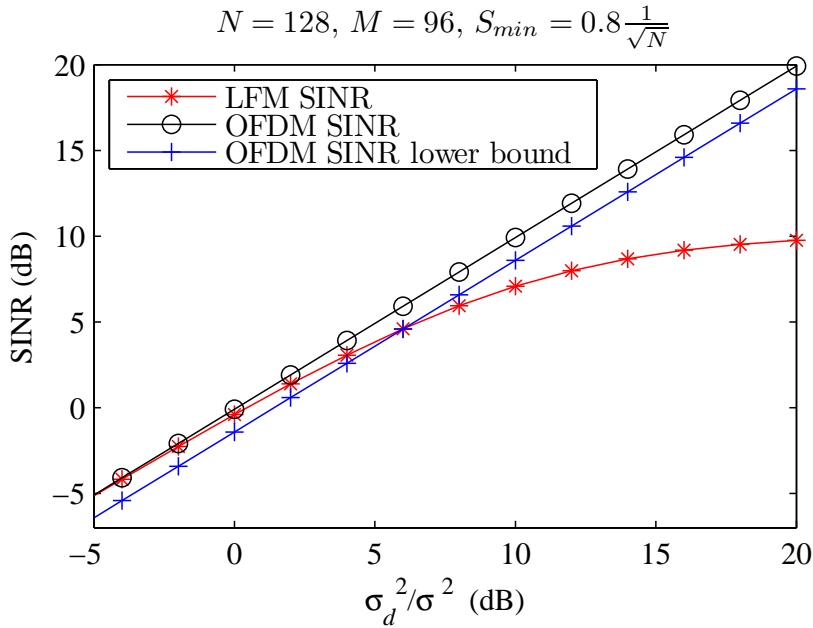}
\end{center}
\caption{The mean SINR comparison using an LFM pulse and a designed OFDM
pulse.}\label{SINRfig}
\end{figure}

In Fig. \ref{SINRfig}, we plot the SINRs when using an LFM pulse as
(\ref{SINRLFMmin}) as well as the SINRs and the lower bounds using the above
designed OFDM pulse with $S_{min}=0.8\frac{1}{\sqrt{N}}$ as (\ref{SINROFDMmin})
versus $\frac{\sigma_d^2}{\sigma^2}$. The curves denote that the SINR lower
bounds using the OFDM pulse are insignificantly smaller than the SINRs using
the LFM pulse for $\frac{\sigma_d^2}{\sigma^2}<6$ dB. However, the SINR lower
bounds using the OFDM pulse are larger than the SINRs using the LFM pulse for
$\frac{\sigma_d^2}{\sigma^2}>6$ dB. Moreover, the advantage of the SINR lower
bounds by using the OFDM pulse is more obvious when
$\frac{\sigma_d^2}{\sigma^2}$ is larger. Furthermore, the true SINRs using the
OFDM pulse are about $1.4$ dB larger than their lower bounds, never smaller
than the SINRs using the LFM pulse for small $\frac{\sigma_d^2}{\sigma^2}$, and
obviously larger than the SINRs using the LFM pulse for
$\frac{\sigma_d^2}{\sigma^2}>0$ dB. These results indicate that the range
reconstruction SNR degradation using a designed OFDM pulse is insignificant,
and the advantage by using a designed OFDM pulse is more significant when noise
power $\sigma^2$ becomes smaller.

\subsection{Performance of the SAR imaging}
\begin{figure}[t]
\begin{center}
\includegraphics[width=0.5\columnwidth,draft=false]{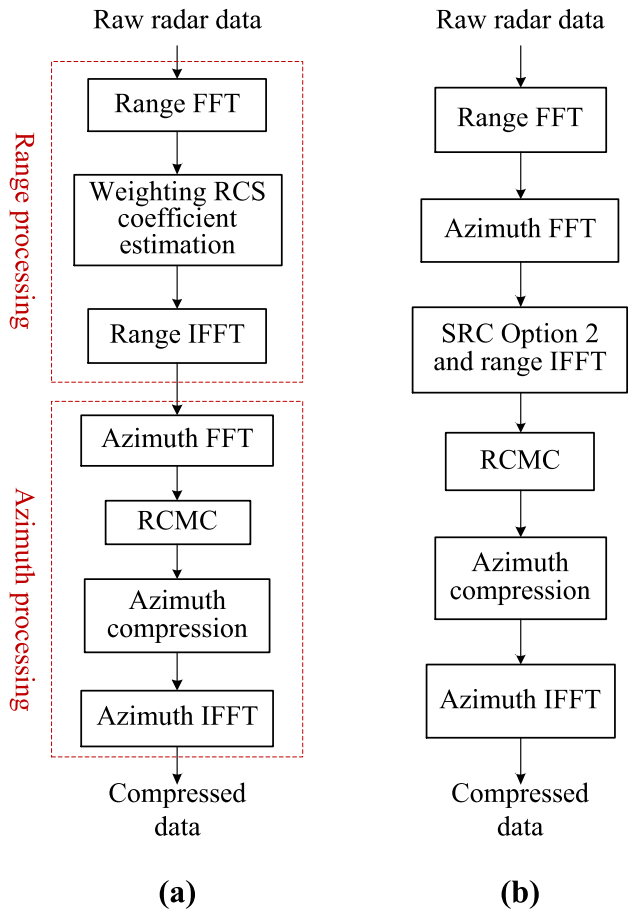}
\end{center}
\caption{Block diagram of SAR imaging processing: (a) CP based OFDM SAR; (b)
LFM SAR.}\label{LFM_Block}
\end{figure}

In this subsection, we present some simulations and discussions for the
proposed CP based arbitrary OFDM pulse length range reconstruction for SAR
imaging. The azimuth processing is similar to the conventional stripmap SAR
imaging \cite{Soumekh1999Synthetic}, and a fixed value of $R_c$ located at the
center of the range swath is set as the reference range cell for azimuth
processing as what is commonly done in SAR image simulations. For comparison,
we also consider the range Doppler algorithm (RDA) using LFM
signals\footnote{Since the performance of random noise SAR is similar to LFM
SAR, we do not present any simulation results of random noise SAR here. For
more comparisons between OFDM SAR imaging, LFM SAR imaging, and random noise
SAR imaging, we refer to \cite{TxzOFDMSAR}.} as shown in the block diagram of
Fig. \ref{LFM_Block}. In Fig. \ref{LFM_Block} (b), the secondary range
compression (SRC) is implemented in the range and azimuth frequency domain, the
same as the Option 2 in \cite[Ch. 6.2]{Soumekh1999Synthetic}.

\begin{figure}[t]%
\centering \subfigure[]{
\label{range_profile} 
\includegraphics[width=5.3in]{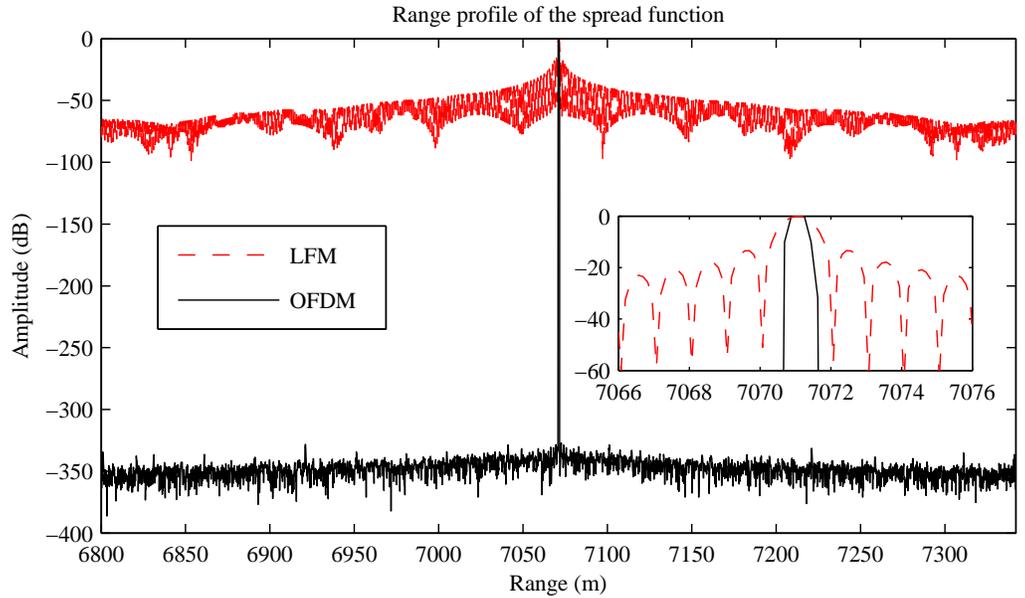}}\\
\hspace{0.04in} \subfigure[]{ \label{azimuth_profile} 
\includegraphics[width=5.3in]{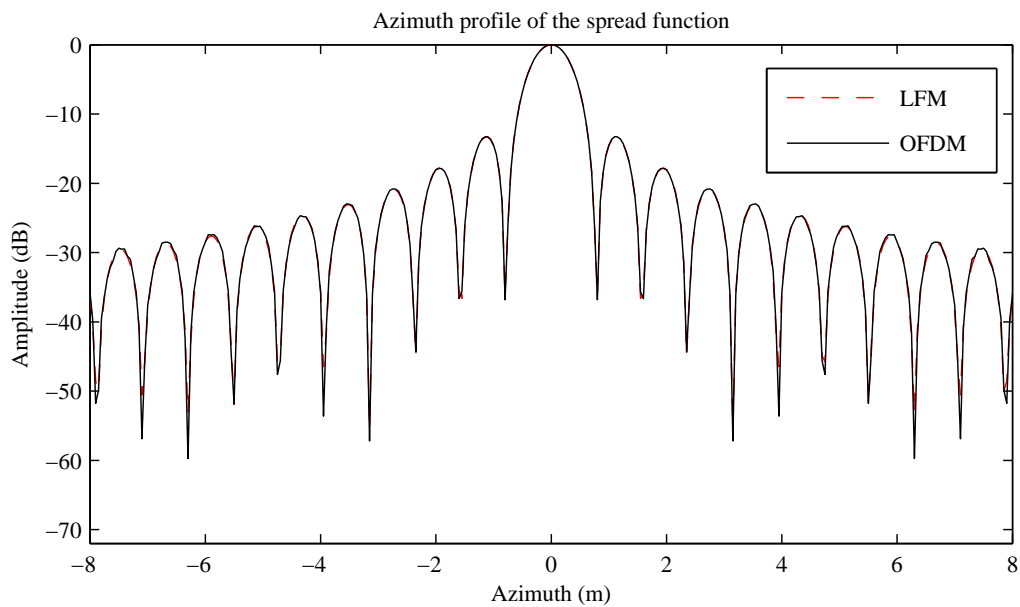}}
\caption{Profiles of a point spread function: (a) range profiles; (b) azimuth
profiles.}
\label{Profiles} 
\end{figure}

The simulation parameters are set as in a typical SAR system: PRF = $800$ Hz,
the bandwidth is $B = 150$ MHz, the antenna length is $L_a=1$ m, the carrier
frequency $f_c=9$ GHz, the synthetic aperture time is $T_a=1$ sec, the
effective radar platform velocity is $v_p=150$ m/sec, the platform height of
the antenna is $H_p=5$ km, the slant range swath center is $R_c=5\sqrt{2}$ km,
the sampling frequency $f_s=150$ MHz.

Firstly, the normalized range profiles and azimuth profiles of a point spread
function are shown in Fig. \ref{Profiles}. It can be seen that the range
sidelobes are much lower for the OFDM signal than those of the LFM signal. And
the azimuth profiles of the point spread function are similar for these two
signals.

\begin{figure}[b]%
\centering \subfigure[]{
\label{RangeLine:a} 
\includegraphics[width=3in]{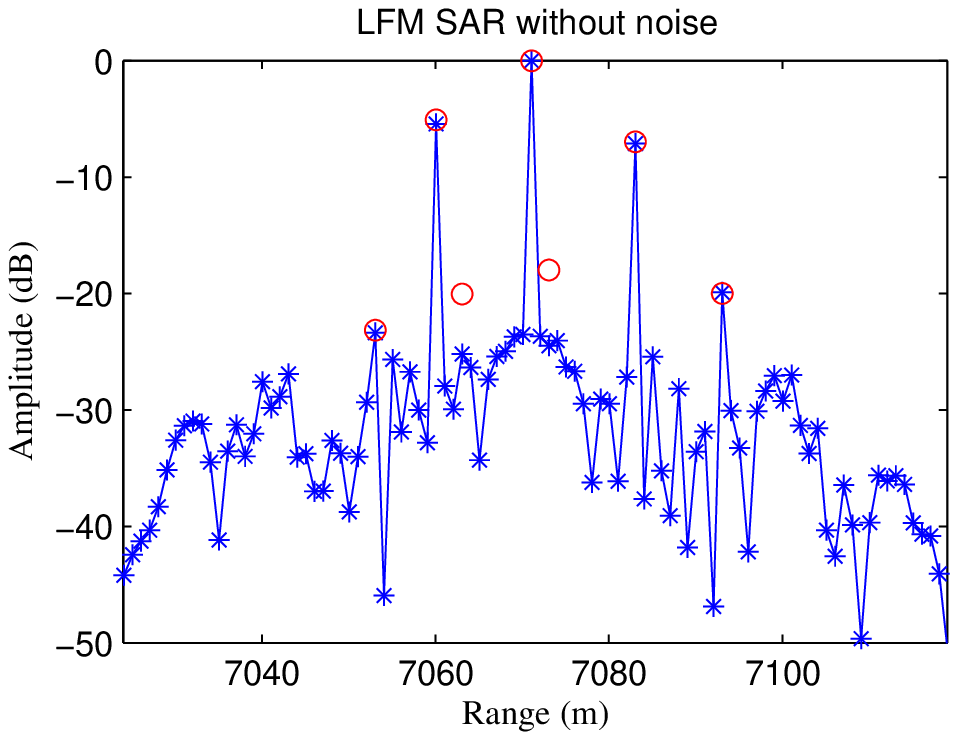}}
\hspace{0.04in} \subfigure[]{
\label{RangeLine:b} 
\includegraphics[width=3in]{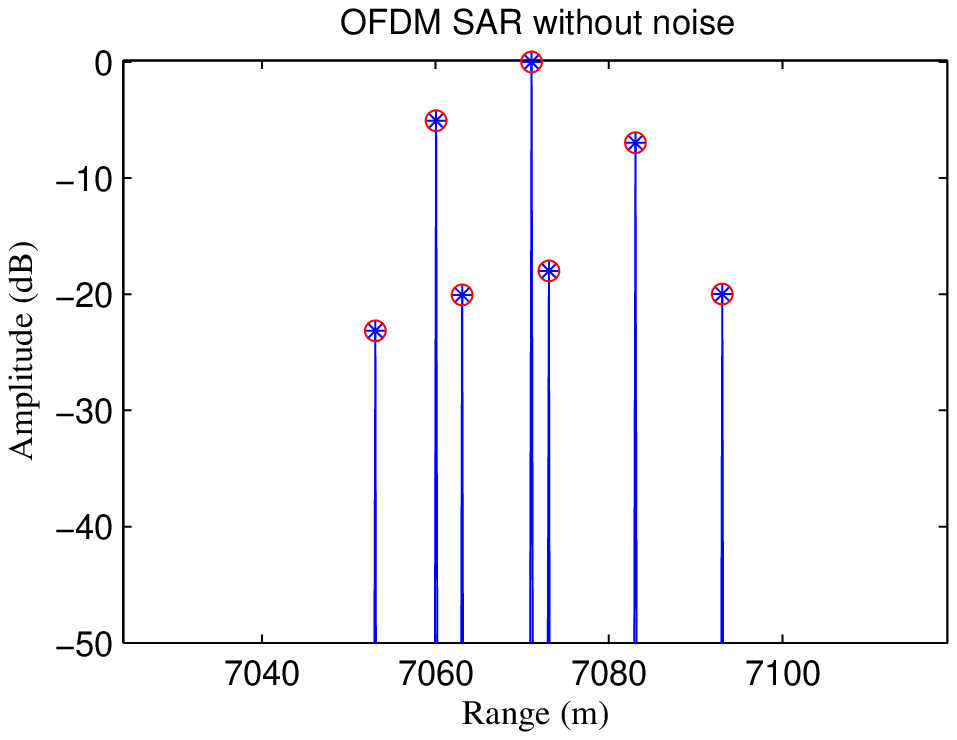}}\\
\subfigure[]{
\label{RangeLine:c} 
\includegraphics[width=3in]{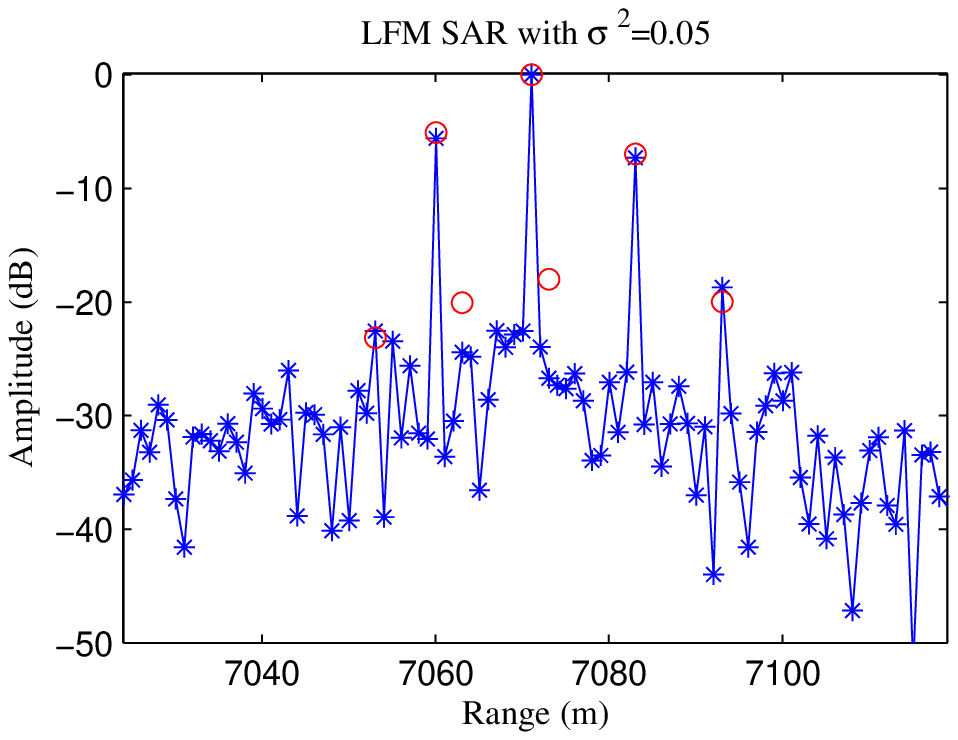}}
\hspace{0.04in} \subfigure[]{
\label{RangeLine:d} 
\includegraphics[width=3in]{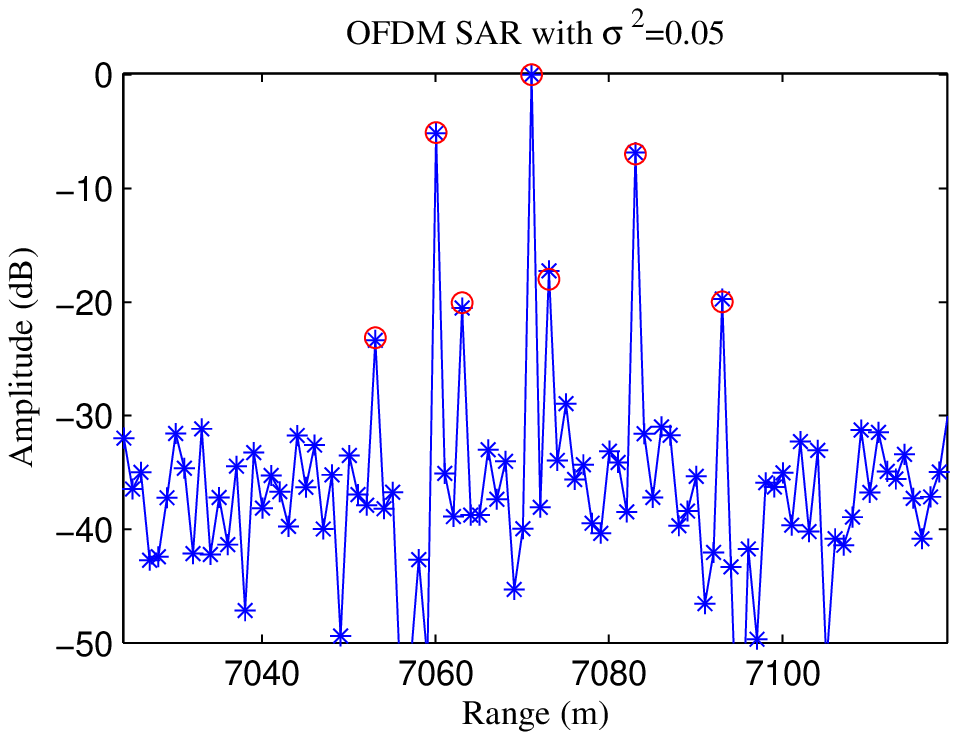}}\\
\subfigure[]{
\label{RangeLine:e} 
\includegraphics[width=3in]{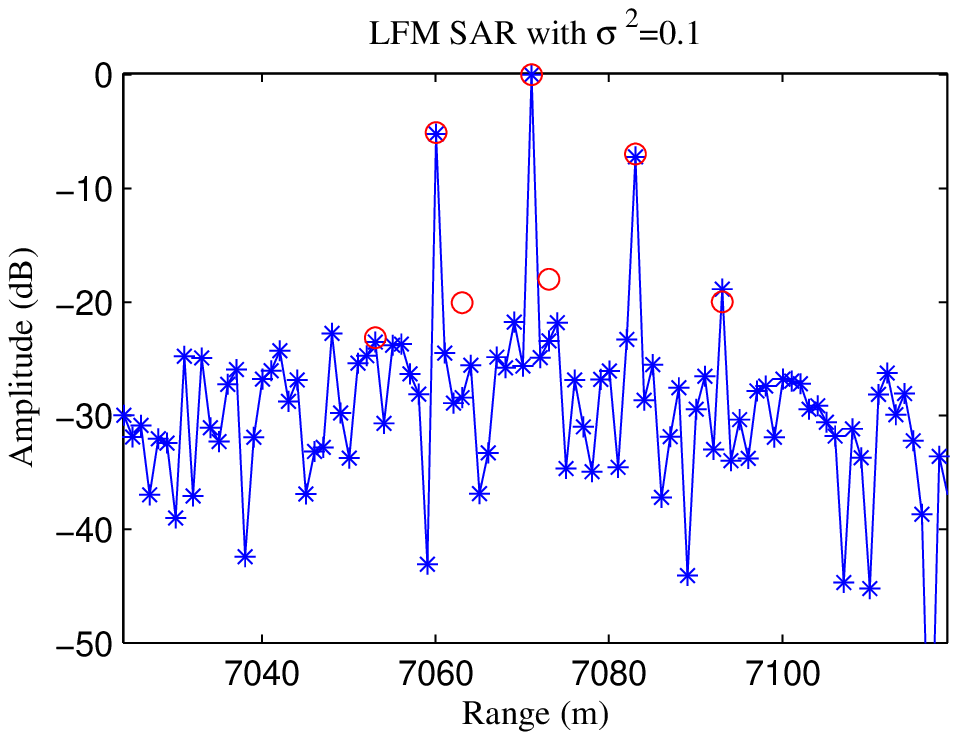}}
\hspace{0.04in} \subfigure[]{
\label{RangeLine:f} 
\includegraphics[width=3in]{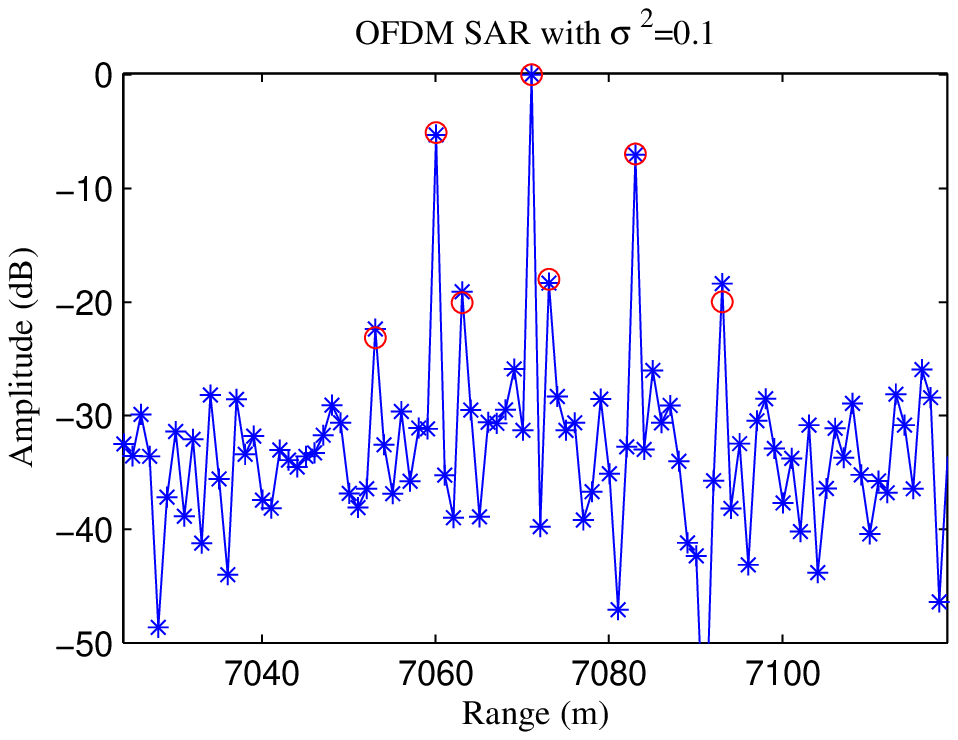}}
\caption{A range line imaging results. Red circles denote the real target
amplitudes, blue asterisks denote the imaging results. (a) LFM SAR without
noise; (b) OFDM SAR without noise; (c) LFM SAR with noise of variance
$\sigma^2=0.05$; (d) OFDM SAR with noise of variance $\sigma^2=0.05$; (e) LFM
SAR with noise of variance $\sigma^2=0.1$; (f) OFDM SAR with noise of variance
$\sigma^2=0.1$.}\label{RangeLine}
\end{figure}

We also consider a single range line (a cross range) with $M=10000$ range cells
in a $10$ km wide swath, and targets (non-zero RCS coefficients) are included
in $7$ range cells located from $7050$ m to $7100$ m, the amplitudes are
randomly generated and shown as the red circles in Fig. \ref{RangeLine}, and
the RCS coefficients of the other range cells are set to be zero (for a better
display, only a segment of the swath is indicated in Fig. \ref{RangeLine}). In
this simulation, we use a designed OFDM pulse with PAPR $= 1.93$ dB, $\xi=
-0.14$ dB and time duration\footnote{For the algorithm in \cite{TxzOFDMSAR}, by
setting $N=M$, the OFDM pulse time duration with sufficient length CP is at
least $T+T_{GI}=\frac{10000}{150}+\frac{9999}{150}\ \mu \textrm{s}\approx
133.3\ \mu$s as mentioned in Section \ref{Problem_For}.} $T-T_{GI}=5\ \mu$s,
which is independent of the swath width. For $T_{GI}=\frac{M-1}{f_s}$,
$N=Tf_s=10749$. The transmitted LFM pulse duration is also $5\ \mu$s. The
normalized imaging results are shown as the blue asterisks in Fig.
\ref{RangeLine}. The imaging results without noise are shown in Fig.
\ref{RangeLine:a} and Fig. \ref{RangeLine:b}. Since  there is no IRCI between
different range cells, the results indicate that the OFDM SAR imaging is
precise as shown in Fig. \ref{RangeLine:b}. However, because of  the influence
of range sidelobes of the LFM signal, some weak targets, for example, those
located at $7063$ m and $7073$ m, are submerged by the interference from the
nearby targets and thus can not be imaged correctly as shown in Fig.
\ref{RangeLine:a}. We also give the imaging results of LFM SAR and OFDM SAR in
Fig. \ref{RangeLine:c} and Fig. \ref{RangeLine:d}, respectively, when the noise
power of the raw radar data is $\sigma^2=0.05$, and in Fig. \ref{RangeLine:e}
and Fig. \ref{RangeLine:f}, respectively, when $\sigma^2=0.1$. These results
can also indicate the better performance of the proposed OFDM SAR. The
performance advantage of the OFDM SAR is more obvious for a smaller noise
power, for example, when $\sigma^2=0.05$, which is consistent with the results
in Fig. \ref{SINRfig}. Note that, for a better display and recognizability, we
consider that only $7$ range cells in the swath contain targets. In a practical
SAR imaging, much more targets (non-zero RCS coefficients) are included and
then the IRCI of LFM (or random noise) SAR will be more serious. Thus, the
performance advantage of the OFDM SAR over LFM or random noise SAR will be more
obvious because of its IRCI free range reconstruction.

\section{Conclusion}\label{Conclusion}
In this paper, we proposed a novel sufficient CP based OFDM SAR imaging
algorithm with arbitrary pulse length that is independent of a swath width by
using our newly proposed and designed OFDM pulses. This OFDM SAR imaging
algorithm can provide the advantage of IRCI free range reconstruction and avoid
the energy redundancy. We first established the arbitrary pulse length OFDM SAR
imaging system model and then derived the range reconstruction algorithm with
free IRCI. We also analyzed the SINR after the range reconstruction and
compared it with that using LFM signals. By considering the PAPR of a
transmitted OFDM pulse and the SNR degradation of the range reconstruction, we
proposed a novel OFDM pulse design method. We finally gave some simulations to
demonstrate the performance of the proposed OFDM pulse design method. By
comparing with the RDA SAR imaging using LFM signals, we provided some
simulations to illustrate the advantage, such as higher SINR after the range
reconstruction, of the proposed arbitrary pulse length OFDM SAR imaging
algorithm. The main contributions of this paper
 can be summarized as:
\begin{itemize}
  \item
When a sufficient CP length is at least $M-1$, where $M$ is the number of range
cells within a swath, an OFDM sequence of length $N$, $\bs\in
\mathbb{C}^{N\times 1}$, with at least $M-1$ consecutive zero elements in the
head part is generated by an OFDM pulse design method and thus, the transmitted
OFDM sequence is $\bs_t\in \mathbb{C}^{(N-M+1)\times 1}$ of length $N+M-1$.
  \item
With our proposed OFDM sequence/pulse design, a transmitted OFDM pulse length
can be arbitrary and independent of a swath width, which is critical in wide
swath IRCI free SAR imaging applications.
  \item
With a designed OFDM pulse, no CP in the transmitted sequence needs to be
removed in the receiver. Thus, the transmitted energy redundancy can be
 avoided.
  \item
The proposed SAR imaging algorithm may cause some SNR degradation. However, the
degradation is insignificant according to our simulations. Comparing with LFM
SAR, the performance advantage of the OFDM SAR is more obvious for a smaller
noise power. Moreover, with our proposed OFDM pulse design method, a better
OFDM sequence with a lower PAPR can be generated by setting a larger maximum
iteration number $Q$, and the SNR degradation by using this OFDM sequence
becomes less.

\end{itemize}


\clearpage
\bibliographystyle{IEEEtran}
\bibliography{IEEEabrv,MyReference}
\end{document}